\documentstyle[preprint,aps,graphicx]{revtex}

\def\beq{\begin{equation}}
\def\eeq{\end{equation}}
\def\beqa{\begin{eqnarray}}
\def\eeqa{\end{eqnarray}}

\def\za{\alpha}
\def\zb{\beta}
\def\lsim{\mathrel{\raise.3ex\hbox{$<$\kern-.75em\lower1ex\hbox{$\sim$}}} }
\def\gsim{\mathrel{\raise.3ex\hbox{$>$\kern-.75em\lower1ex\hbox{$\sim$}}} }

\begin{document}
\draft
\preprint{\tighten{\vbox{\hbox{IPAS-HEP-k007}\hbox{Jan 2001}\hbox{rev. Apr 2001}}}}

\title{$\mu \to e\,\gamma$ from Supersymmetry without R-parity}
\author{\bf Kingman Cheung$^{1}$ and Otto C. W. Kong$^2$}
\address{$^1$ National Center for Theoretical Science, NTHU, Hsinchu, Taiwan
\\
$^2$ Institute of Physics, Academia Sinica, Nankang, Taipei, Taiwan 11529}
\maketitle

\begin{abstract}
Starting from the new sources of $LL$- and $LR$- scalar (slepton) mixings due to 
the R-parity violation, we discuss the structure of lepton-flavor violation
focusing on the radiative decay of a muon into an electron.  Using an optimal
parametrization, we give the general  formulas for the one-loop contributions,
from which we discuss all combinations of R-parity violating parameters that 
possibly have a substantial contribution to the $\mu \to e\, \gamma$ decay width. 
An exact numerical study is performed to obtain explicit bounds on the
parameters under the present experimental limit. The most interesting one
involves a combination of bilinear and trilinear couplings.
\end{abstract}


\newpage

\section{Introduction}
The standard model (SM) is built upon the fact that neutrinos have no 
right-handed components and are massless. Thus, lepton flavors are conserved.  
So far, experiments have not yet observed any lepton-flavor violation (LFV).
Nevertheless, the evidence of small neutrino masses (implied from
oscillation analyses of solar and atmospheric neutrino data \cite{osc}) 
implies a small amount of LFV. Many new physics models beyond the SM 
predict a certain amount of LFV, in relation to neutrino mass generation
or otherwise. An important criterion for a viable model is that it should give
an acceptable neutrino mass spectrum while staying within the experimental 
limits of LFV. Among the latter processes, the radiative decay of muon,
$\mu \to e\,\gamma$, provides the most stringent bound \cite{ko}. 
The current limit is \cite{exp}
\[
B(\mu \to e\, \gamma) < 1.2\times 10^{-11} \; .
\]
A proposal is underway, which can improve this limit by 
three orders of magnitude\cite{nexp}.
 
In the Lagrangian of the minimal supersymmetric standard model (MSSM),
apart from the soft supersymmetry (SUSY) breaking part, there are no 
lepton-flavor violating terms that would induce lepton-flavor changing
processes such as $\mu \to e\,\gamma$. It is, however, well known that 
soft SUSY breaking parameters can have all types of flavor-changing terms.
Limits on such parameters from $\mu \to e\,\gamma$ have been extensively
studied \cite{mssm}. On the other hand, flavor-changing soft SUSY breaking 
parameters are often expected to be suppressed by the source of SUSY breaking 
of a more fundamental theory. A particularly interesting 
example is the gauge mediation models \cite{gmsb}. 

Models with ``right-handed neutrinos" provide other interesting alternatives 
with potentially large LFV\cite{gut}. It is no surprise that issues of LFV are 
often tied up with the generation of neutrino masses. Within the MSSM, 
apart from the soft terms, both LFV and neutrino masses are indeed forbidden 
by an {\it ad hoc} discrete symmetry --- R parity. Note that the soft terms 
by themselves still conserve total lepton number, and hence do not generate 
neutrino masses.

In the supersymmetric standard model without R-parity imposed, 
there is however another important source of LFV. A major part of this comes 
simply from the R-parity violating (RPV) terms in the superpotential. 
We would like to emphasize that the latter is independent of SUSY breaking 
and mediation mechanisms. In fact, the only theoretical means to suppress the 
flavor-changing effects from RPV couplings would be a generic flavor theory, 
or some {\it ad hoc} symmetries such as R-parity itself. Hence, it is
important to obtain all the available experimental bounds on the RPV
couplings. We show that the experimental limit on the radiative decay 
of muon, $\mu \to e\,\gamma$, provides one of the most stringent constraints on 
the RPV couplings. An improvement of a few orders of magnitude in the experiment
may, in fact, discover SUSY as well as the LFV, if the model under 
consideration does explain the nature.

We first explain a new contribution to the $LR$-slepton mass mixings
involving the bilinear $\mu_i$ and trilinear $\lambda_{ijk}$ RPV superpotential
parameters. Such an important interplay between the bilinear and trilinear 
couplings was first discussed in Refs.\cite{as1,as5} in the context of neutrino mass. 
There is also an analogous contribution to the $LR$-squark mixings, with implications
to 1-loop neutrino masses\cite{as5} and fermion electric and magnetic dipole moments
\cite{as4,as6,cch1,as11}.  With nonzero $\mu_k$ and $\lambda_{kij}$, the (off-diagonal) 
$LR$-mixing involving the $i,j$ generations is proportional to 
$\mu_k^* \,\lambda_{kij} \quad \mbox{(sum over $k$)}$. With no particular arguments 
or theories to enforce $\lambda_{kij}$  to be diagonal in the last two indices, 
the LFV proportional to $\mu_k^* \,\lambda_{kij}$ turns out to be quite natural. 
The combination $\mu_k^* \,\lambda_{kij}$ conserves the overall lepton number 
but violates lepton flavor. There is another similar contribution to slepton
masses in the $LL$ mixing part in the form of $\mu_i^* \,\mu_j$. 
So far as we know, the implication of such terms in the slepton mass matrix
is discussed here for the first time. The structure also illustrates very clearly
the basics of the LFV structure, which comes into $\mu \to e\,\gamma$, for example,
in other ways too, as we will see below. 

In this article, we presented a systematic analysis of the $\mu \to e\,\gamma$ 
emphasizing the  new RPV contributions.  We work within the framework 
of the single-VEV parametrization (SVP)\cite{k,as8}, 
which is an optimal choice of flavor basis that helps to guarantee 
a consistent and unambiguous treatment of all kinds of admissible RPV couplings 
and to maintain a simple structure for RPV effects on tree-level mass matrices 
for all states including scalars and fermions\cite{smass}. Note that under the SVP,
the above-mentioned terms with LFV represent the full RPV contribution to
the corresponding entries in the (charged) slepton mass matrix. Obviously, the
$\mu_i^* \,\mu_j$ combinations contribute equally to the neutral scalar
(``sneutrino") masses. There are other interesting RPV contributions that
mix the sleptons with the other scalars (``Higgses"). Among the latter are the 
interesting contributions from the soft $B_i$ parameters. They come into
$\mu \to e\,\gamma$ at a different level.

The present work differs from previous ones on the topic
\cite{DW,CD,CH,cch2,other} in a few important ways.  Ours is
based on the most generic theory of SUSY without R-parity, {\it i.e.}, 
no assumptions on the form of R-parity violation, while most of the
previous authors worked in various limited RPV scenarios. The only exception is
the recent paper of Ref.\cite{cch2}, which basically works under the same framework
as ours. They include the $LR$-slepton mixing contribution, following the suggestion of Ref.\cite{as4}, where the closely related subject of fermion electric dipole 
moments is discussed. However, the results presented are not as complete and 
systematic as our treatment here. In particular, there is no discussion of the LFV 
from the bilinear RPV parameters only.  In addition, their bounds on 
$\mu_k^* \,\lambda_{kij}$ have a $\cos\!\zb$ dependence, which we disagree with.
Other than that, the above new RPV contributions have not been studied before.

The organization of the paper is as follows. In Sec. II, we give the
general formulas in the basis of mass eigenstates. The focus is on  the 
$\ell^{\!\!\mbox{ -}}_j  \to \ell^{\!\!\mbox{ -}}_i \, \gamma$
amplitude from 1-loop diagrams without colored intermediate states.
Though the small deviations of the external
physical charged lepton mass eigenstates ($\ell_i^\pm$'s) from the electroweak
states $l^\pm_i$'s are neglected, full attention is given to such deviations 
for the particles running in the loop when we analyze the diagrams. Likewise,
exact mass eigenstates are always used in the numerical calculations of the
1-loop amplitudes. The formulas can be easily adapted for the 
radiative $\tau$ decay with simple substitutions.
Guided by the formulas and results from our exact numerical calculation, we analyze 
in detail the important pieces of 1-loop contributions in Sec.~III. 
Numerical results are presented and discussed in Sec.~IV; after 
which we conclude the article in Sec.~V. For the readers who are more interested
in the major features of the result, rather than the sophisticated details
on how each RPV parameter comes into play in the process, Sec.~III can simply be
skipped. The section also serves as an analytical confirmation of the correctness
of the numerical results, for the more serious readers.

\section{Calculation in mass basis}
Here we calculate the decay rate of $\mu \to e\,\gamma$ through colorless 1-loop 
diagrams using exact mass eigenstates for particles appearing inside the loop. 
However, as stated above, for the external legs to the loop, we use only electroweak 
states (${l}_i^{\mp}$'s). Each of the latter has a minimal deviation from 
the corresponding $L$- or $R$-handed component of the physical charged lepton 
states as a result of the corresponding small RPV $\mu_i$ terms\cite{k}. 
 
\subsection{Model background}
We work in the framework of the generic supersymmetric standard model\cite{as8,as12}, 
known as SUSY without R-parity. Here one could simply give up the notion of 
R-parity; however, the latter terminology is convenient for comparison 
with the more popular MSSM, as well as other limited studies of R-parity
violations. We adopt exactly the same notation as in Refs.\cite{as6,as8}, to 
which readers are referred for more details. Let us summarize some of the 
useful notation here, starting from the generic superpotential
\[
W = \varepsilon_{ab}\left[ \mu_{\alpha}  \hat{H}_u^a \hat{L}_{\alpha}^b 
+ y_{ik}^u \hat{Q}_i^a   \hat{H}_{u}^b \hat{U}_k^{\scriptscriptstyle C}
+ \lambda_{\alpha jk}^{\!\prime}  \hat{L}_{\alpha}^a \hat{Q}_j^b
\hat{D}_k^{\scriptscriptstyle C} + 
\frac{1}{2}\, \lambda_{\alpha \beta k}  \hat{L}_{\alpha}^a  
 \hat{L}_{\beta}^b \hat{E}_k^{\scriptscriptstyle C} \right] + 
\frac{1}{2}\, \lambda_{ijk}^{\!\prime\prime}  
\hat{U}_i^{\scriptscriptstyle C} \hat{D}_j^{\scriptscriptstyle C}  
\hat{D}_k^{\scriptscriptstyle C}  \; 
\]
(with $\alpha,\beta$ going from $0$ to $3$, $i,j,k$ from 1 to 3, and 
$\epsilon_{\scriptscriptstyle 12}=-\epsilon_{\scriptscriptstyle 21}=1$). 
The essential features of the SVP adopted include
$\left\langle \hat{L}_i \right\rangle \equiv 0$ and the identification of 
$\hat{L}_{0}$ as $\hat{H}_d$; related to that is the 
fact that $\lambda_{{\scriptscriptstyle 0}ii}\equiv 
{y_{\!\scriptscriptstyle e_i}}$ is the diagonal ``leptonic Yukawa" couplings, 
while $\lambda_{{\scriptscriptstyle 0}ij}\equiv 0$ for $i\ne j$.

(i) We have five (color-singlet) charged fermions from
$R$- and $L$-handed mass eigenstates written, respectively, as
$(\, \chi_{{\!\scriptscriptstyle +}n} \,)= 
\mbox{\boldmath $V$}^{\scriptscriptstyle T} \,
[ -i\tilde{W}^{\scriptscriptstyle +},
\tilde{h}_{\!\scriptscriptstyle u}^{\!\scriptscriptstyle +},
{l}_{\scriptscriptstyle 1}^{\!\scriptscriptstyle +},
{l}_{\scriptscriptstyle 2}^{\!\scriptscriptstyle +},
{l}_{\scriptscriptstyle 3}^{\!\scriptscriptstyle +}\,]^{\scriptscriptstyle T}$
and
$(\, \chi_{\!\!\!\mbox{ -}n} \,)= \mbox{\boldmath $U$}^{\dag} \,
[-i\tilde{W}^{\!\!\mbox{ -}},
{l}_{\scriptscriptstyle 0}^{\!\!\mbox{ -}},
{l}_{\scriptscriptstyle 1}^{\!\!\mbox{ -}},
{l}_{\scriptscriptstyle 2}^{\!\!\mbox{ -}},
{l}_{\scriptscriptstyle 3}^{\!\!\mbox{ -}}\,]^{\scriptscriptstyle T}$, 
where the notation for the electroweak states is quite obvious. So,
$\mbox{\boldmath $V$}^\dag {\mathcal{M}_{\scriptscriptstyle C}} \,
\mbox{\boldmath $U$} = \mbox{diag} 
\{ {M}_{\!\scriptscriptstyle \chi^{\mbox{-}}_n} \} \equiv 
\mbox{diag} 
\{ {M}_{c {\scriptscriptstyle 1}}, {M}_{c {\scriptscriptstyle 2}},
m_e, m_\mu, m_\tau \}$. The first two mass eigenvalues are the chargino masses.
Note that under the SVP, the only RPV parameters going into the tree level 
mass matrix ${\mathcal{M}_{\scriptscriptstyle C}}$ are the three $\mu_i$'s.

(ii) The neutral fermion mass matrix $\cal{M_{\scriptscriptstyle N}}$ 
is written  in the  basis 
$(-i\tilde{B}, -i\tilde{W}, 
\tilde{h}_{\!\scriptscriptstyle u}^{\!\scriptscriptstyle 0}\,, 
\tilde{h}_{\!\scriptscriptstyle d}^{\!\scriptscriptstyle 0}\,, 
{l}_{\scriptscriptstyle 1}^{\scriptscriptstyle 0}\,,
{l}_{\scriptscriptstyle 2}^{\scriptscriptstyle 0}\,,
{l}_{\scriptscriptstyle 3}^{\scriptscriptstyle 0}\,)$,
where 
$\tilde{h}_{\!\scriptscriptstyle d}^{\!\scriptscriptstyle 0}
\equiv {l}_{\scriptscriptstyle 0}^{\scriptscriptstyle 0}$.
The symmetric, but generically non-hermitian, mass matrix is diagonalized by a 
unitary matrix {\boldmath $X$} such that
$\mbox{\boldmath $X$}^{\!\scriptscriptstyle  T} 
{\cal M_{\scriptscriptstyle N}}\mbox{\boldmath $X$} =
\mbox{diag} \{ {M}_{\!\scriptscriptstyle \chi^0_{n}} \}$,
with $n = 1$ to $4$ being the heavy states (neutralinos) and  $n = 5$ to $7$ 
the physical neutrino states at tree-level. 

(iii) There are $1+4+3$ charged scalars that contribute one
unphysical Goldstone state after diagonalization. We use the 
basis 	$\{\, h_u^{{\!\scriptscriptstyle +}\dag}, 
\tilde{l}_{\scriptscriptstyle 0}^{\!\!\mbox{ -}},
\tilde{l}_{\scriptscriptstyle 1}^{\!\!\mbox{ -}},
\tilde{l}_{\scriptscriptstyle 2}^{\!\!\mbox{ -}},
\tilde{l}_{\scriptscriptstyle 3}^{\!\!\mbox{ -}},
\tilde{l}_{\scriptscriptstyle 1}^{{\!\scriptscriptstyle +}\dag},
\tilde{l}_{\scriptscriptstyle 2}^{{\!\scriptscriptstyle +}\dag},
\tilde{l}_{\scriptscriptstyle 3}^{{\!\scriptscriptstyle +}\dag}
\,\}$
to write the mass-squared matrix as follows :
\begin{equation}
 \label{ME}
{\cal M}_{\!\scriptscriptstyle {E}}^2 =
\left( \begin{array}{ccc}
\widetilde{\cal M}_{\!\scriptscriptstyle H\!u}^2 &
\widetilde{\cal M}_{\!\scriptscriptstyle LH}^{2\dag}  & 
\widetilde{\cal M}_{\!\scriptscriptstyle RH}^{2\dag}
\\
\widetilde{\cal M}_{\!\scriptscriptstyle LH}^2 & 
\widetilde{\cal M}_{\!\scriptscriptstyle LL}^{2} & 
\widetilde{\cal M}_{\!\scriptscriptstyle RL}^{2\dag} 
\\
\widetilde{\cal M}_{\!\scriptscriptstyle RH}^2 &
\widetilde{\cal M}_{\!\scriptscriptstyle RL}^{2} & 
\widetilde{\cal M}_{\!\scriptscriptstyle RR}^2  
\end{array} \right) \; ;
\end{equation}
where
\begin{eqnarray}
\widetilde{\cal M}_{\!\scriptscriptstyle H\!u}^2 &=&
\tilde{m}_{\!\scriptscriptstyle H_{\!\scriptscriptstyle u}}^2
+ \mu_{\!\scriptscriptstyle \za}^* \mu_{\scriptscriptstyle \za}
+ M_{\!\scriptscriptstyle Z}^2\, \cos\!2 \beta 
\left[ \,\frac{1}{2} - \sin\!^2\theta_{\!\scriptscriptstyle W}\right]
\nonumber \\
&+&  M_{\!\scriptscriptstyle Z}^2\,  \sin\!^2 \beta \;
[1 - \sin\!^2 \theta_{\!\scriptscriptstyle W}]  \; ,
\nonumber \\
\widetilde{\cal M}_{\!\scriptscriptstyle LL}^2 &=&
\tilde{m}_{\!\scriptscriptstyle {L}}^2 +
m_{\!\scriptscriptstyle L}^\dag m_{\!\scriptscriptstyle L}
+ (\mu_{\!\scriptscriptstyle \za}^* \mu_{\scriptscriptstyle \zb})
+ M_{\!\scriptscriptstyle Z}^2\, \cos\!2 \beta 
\left[ -\frac{1}{2} +  \sin\!^2 \theta_{\!\scriptscriptstyle W}\right] 
\; {\large 1}_{4\times 4} \; ,
\nonumber \\
&+& \left( \begin{array}{cc}
 M_{\!\scriptscriptstyle Z}^2\,  \cos\!^2 \beta \;
[1 - \sin\!^2 \theta_{\!\scriptscriptstyle W}] 
& \quad 0_{\scriptscriptstyle 1 \times 3} \quad \\
0_{\scriptscriptstyle 3 \times 1} & 0_{\scriptscriptstyle 3 \times 3}  
\end{array} \right) \; ,
\nonumber \\
\widetilde{\cal M}_{\!\scriptscriptstyle RR}^2 &=&
\tilde{m}_{\!\scriptscriptstyle {E}}^2 +
m_{\!\scriptscriptstyle E} m_{\!\scriptscriptstyle E}^\dag
+ M_{\!\scriptscriptstyle Z}^2\, \cos\!2 \beta 
\left[  - \sin\!^2 \theta_{\!\scriptscriptstyle W}\right] \; 
{\large 1}_{3\times3} \; ;
\end{eqnarray}
and
\begin{eqnarray}
\widetilde{\cal M}_{\!\scriptscriptstyle LH}^2
&=& (B_{\za}^*)  
+ \left( \begin{array}{c} 
{1 \over 2} \,
M_{\!\scriptscriptstyle Z}^2\,  \sin\!2 \beta \;
[1 - \sin\!^2 \theta_{\!\scriptscriptstyle W}]  \\
0_{\scriptscriptstyle 3 \times 1} 
\end{array} \right)
\; ,
\nonumber \\
\widetilde{\cal M}_{\!\scriptscriptstyle RH}^2
&=&  -\,(\, \mu_i^*\lambda_{i{\scriptscriptstyle 0}k}\, ) \; 
\frac{v_{\scriptscriptstyle 0}}{\sqrt{2}} \; 
= (\, \mu_k^* \, m_k \, ) \hspace*{1in} \mbox{ (no sum over $k$)} \quad \; ,
\nonumber \\
(\widetilde{\cal M}_{\!\scriptscriptstyle RL}^{2})^{\scriptscriptstyle T} 
&=& \left(\begin{array}{c} 
0  \\   A^{\!{\scriptscriptstyle E}} 
\end{array}\right)
 \frac{v_{\scriptscriptstyle 0}}{\sqrt{2}}
 - (\, \mu_{\scriptscriptstyle \za}^*\lambda_{{\scriptscriptstyle \za\zb}k}\, ) \; 
\frac{v_{\scriptscriptstyle u}}{\sqrt{2}} \; 
\nonumber \\
&=& [A_e - \mu_{\scriptscriptstyle 0}^* \, \tan\!\beta ] 
\left(\begin{array}{c}  
0  \\   m_{\!{\scriptscriptstyle E}} 
\end{array}\right) \,
+ \frac{\sqrt{2}\, M_{\!\scriptscriptstyle W} \cos\!\beta}
{g_{\scriptscriptstyle 2} } \,
\left(\begin{array}{c} 
0  \\ \delta\! A^{\!{\scriptscriptstyle E}}
\end{array}\right)
-  \left(\begin{array}{c}  
- \mu_{k}^* \, m_k\, \tan\!\beta \\ 
\frac{\sqrt{2}\, M_{\!\scriptscriptstyle W} \sin\!\beta}
{g_{\scriptscriptstyle 2} } \,(\, \mu_i^*\lambda_{ijk}\, ) 
\end{array}\right) \; .
\label{ERL}
\end{eqnarray}
Introducing the diagonalizing matrix ${\cal D}^{l}$, we have
${\cal D}^{l\dag}  {\cal M}_{\!\scriptscriptstyle E}^2 \, {\cal D}^{l}
= \mbox{diag}\{\,M_{\!\scriptscriptstyle \tilde{\ell}_m}^2, m=1\,\mbox{to}\,8\,\}$.
We label the unphysical Goldstone mode by $m=1$. With relatively small
RPV parameters the ${\cal M}_{\!\scriptscriptstyle E}^2$ is predominantly 
diagonal, apart from the mixing between the Higgs bosons ({\it i.e.},
$h_u$ and $h_d \equiv \tilde{l}_{\scriptscriptstyle 0}$) to give the $m=1$
mode. The matrix ${\cal D}^{l}$ may then be naturally organized 
with all diagonal entries being of order one and only
the $12$ and $21$ entries being possibly large (of order one) among the 
off-diagonal ones.

(iv) The neutral scalar mass terms, in terms of the
$(1+4)$ complex scalar fields,  $\phi_n$'s, can be written in two parts
--- a simple $({\cal M}_{\!\scriptscriptstyle {\phi}{\phi}^{\!\dag}}^2)_{mn} \,
\phi_m^\dag \phi_n$ part, and a Majorana-like part in the form 
${1\over 2} \,  ({\cal M}_{\!\scriptscriptstyle {\phi\phi}}^2)_{mn} \,
\phi_m \phi_n + \mbox{h.c.}$. As the neutral scalars are originated
from chiral doublet superfields, the existence of the Majorana-like
part is a direct consequence of the electroweak symmetry
breaking VEVs, hence restricted to the scalars playing the Higgs
role only. They come from the quartic terms of the Higgs fields in
the scalar potential. Using the $\phi_n$ basis 
$(\,{h}_{\!\scriptscriptstyle u}^{{\!\scriptscriptstyle 0}\dag}\,, 
\tilde{l}_{\scriptscriptstyle 0}^{\scriptscriptstyle 0}\,, 
\tilde{l}_{\scriptscriptstyle 1}^{\scriptscriptstyle 0}\,,
\tilde{l}_{\scriptscriptstyle 2}^{\scriptscriptstyle 0}\,,
\tilde{l}_{\scriptscriptstyle 3}^{\scriptscriptstyle 0}\,)$,
we have, explicitly,
\begin{eqnarray}
 \label{Mpp}
{\cal M}_{\!\scriptscriptstyle {\phi\phi}}^2 =
{1\over 2} \, M_{\!\scriptscriptstyle Z}^2\,
\left( \begin{array}{ccc}
 \sin\!^2\! \beta  &  - \cos\!\beta \, \sin\! \beta
& \quad 0_{\scriptscriptstyle 1 \times 3} \\
 - \cos\!\beta \, \sin\! \beta & \cos\!^2\! \beta 
& \quad 0_{\scriptscriptstyle 1 \times 3} \\
0_{\scriptscriptstyle 3 \times 1} & 0_{\scriptscriptstyle 3 \times 1} 
& \quad 0_{\scriptscriptstyle 3 \times 3} 
\end{array} \right) \; ;
\end{eqnarray}
and
\begin{eqnarray}
{\cal M}_{\!\scriptscriptstyle {\phi}{\phi}^{\!\dag}}^2 &=&
\left( \begin{array}{cc}
\tilde{m}_{\!\scriptscriptstyle H_{\!\scriptscriptstyle u}}^2
+ \mu_{\!\scriptscriptstyle \za}^* \mu_{\scriptscriptstyle \za}
+ M_{\!\scriptscriptstyle Z}^2\, \cos\!2 \beta 
\left[-\frac{1}{2}\right]   
& - (B_\za) \\
- (B_\za^*) &
\tilde{m}_{\!\scriptscriptstyle {L}}^2 
+ (\mu_{\!\scriptscriptstyle \za}^* \mu_{\scriptscriptstyle \zb})
+ M_{\!\scriptscriptstyle Z}^2\, \cos\!2 \beta 
\left[ \frac{1}{2}\right]\end{array} \right) 
+  {\cal M}_{\!\scriptscriptstyle {\phi\phi}}^2 \; .
\label{Mp}
\end{eqnarray}
Note that ${\cal M}_{\!\scriptscriptstyle {\phi\phi}}^2$ here is 
real (see Appendix A of Ref.\cite{as5}), while 
${\cal M}_{\!\scriptscriptstyle {\phi}{\phi}^{\!\dag}}^2$ does have complex entries.
Writing the five $\phi_n$'s in terms of their scalar and pseudoscalar parts,
the full $10\times 10$ (real and symmetric) mass-squared matrix for 
the real scalars is then given by
\begin{equation} \label{MSN}
{\cal M}_{\!\scriptscriptstyle S}^2 =
\left( \begin{array}{cc}
{\cal M}_{\!\scriptscriptstyle SS}^2 &
{\cal M}_{\!\scriptscriptstyle SP}^2 \\
({\cal M}_{\!\scriptscriptstyle SP}^{2})^{\!\scriptscriptstyle T} &
{\cal M}_{\!\scriptscriptstyle PP}^2
\end{array} \right) \; ,
\end{equation}
where the scalar, pseudoscalar, and mixing parts are
\begin{eqnarray}
{\cal M}_{\!\scriptscriptstyle SS}^2 &=&
\mbox{Re}({\cal M}_{\!\scriptscriptstyle {\phi}{\phi}^{\!\dag}}^2)
+ {\cal M}_{\!\scriptscriptstyle {\phi\phi}}^2 \; ,
\nonumber \\
{\cal M}_{\!\scriptscriptstyle PP}^2 &=&
\mbox{Re}({\cal M}_{\!\scriptscriptstyle {\phi}{\phi}^{\!\dag}}^2)
- {\cal M}_{\!\scriptscriptstyle {\phi\phi}}^2 \; ,
\nonumber \\
{\cal M}_{\!\scriptscriptstyle SP}^2 &=& - 
\mbox{Im}({\cal M}_{\!\scriptscriptstyle {\phi}{\phi}^{\!\dag}}^2) \; ,
\label{lastsc} 
\end{eqnarray} 
respectively. If $\mbox{Im}({\cal M}_{\!\scriptscriptstyle {\phi}{\phi}^{\!\dag}}^2)$
vanishes, the scalars and pseudoscalars decouple from one another and 
the unphysical Goldstone mode would be found among the latter. Note that 
the $B_\za$ entries may also be considered as a kind of $LR$ mixings.

As a real scalar mass matrix, ${\cal M}_{\!\scriptscriptstyle S}^2$
could be diagonalized by an orthogonal matrix ${\cal D}^{s}$. However, we will
write ${\cal D}^{s}$ as if it is just a unitary matrix. This would be useful
for illustrating some theoretical features in the discussions below. In fact, it helps sometimes to think about the neutral scalars as complex scalars instead of 
in terms of the scalar and pseudoscalar constituents. This is especially true
for the ``sneutrino" parts. Hence, we write
${\cal D}^{s\dag}  {\cal M}_{\!\scriptscriptstyle S}^2 \, 
{\cal D}^{s} = \mbox{diag}\{\,M_{\!\scriptscriptstyle S_m}^2, m=1\,\mbox{to}\,10\,\}$.
Again, it is useful to consider the form of ${\cal D}^{s}$ very close
to the identity matrix, {\it i.e.}, with all diagonal entries being of order one. 
The unphysical Goldstone 
mode has, of course, to be found mainly among the first two
pseudoscalars. The mode is naturally labeled as the $m=6$ mass eigenstate here.
Similar to the previous case, all the off-diagonal entries, except those related 
to mixing of the Higgs bosons ({\it i.e.}, the 12, 21, 67, and 76 entries)
are  expected to be relatively small.

(v) In the diagonalization of the neutral- and charged-scalar mass-squared matrices
given above, one has to impose all the proper tadpole conditions to get the 
correct unphysical Goldstone modes explicitly. Under the SVP, the tadpole 
conditions are (see Appendix A of Ref.\cite{as5} for more details)
\begin{eqnarray}
B_0 \, \cot\!\beta &=& { \left[ 
\tilde{m}_{\!\scriptscriptstyle H_{\!\scriptscriptstyle u}}^2
+ \mu_{\!\scriptscriptstyle \za}^* \mu_{\scriptscriptstyle \za}
+ \frac{1}{8}(g_{\scriptscriptstyle 1}^2  + g_{\scriptscriptstyle 2}^2) 
(v_{\scriptscriptstyle u}^2 -v_{\scriptscriptstyle d}^2)  \right]} \; ,
\label{tpu}\\
B_0 \, \tan\!\beta &=& { \left[ 
\tilde{m}^2_{\!{\scriptscriptstyle L}_{\!{\scriptscriptstyle 00}}}
+ |\mu_{\scriptscriptstyle 0}|^2 + \frac{1}{8}(g_{\scriptscriptstyle 1}^2  
+ g_{\scriptscriptstyle 2}^2) 
(v_{\scriptscriptstyle d}^2 -v_{\scriptscriptstyle u}^2)  \right] }  \; ,
\label{tpd}\\
B_i \, \tan\!\beta &=& 
\tilde{m}^2_{\!{\scriptscriptstyle L}_{\!{\scriptscriptstyle 0}i} }
+ \mu_{\scriptscriptstyle 0}^{*}\, \mu_i \qquad\qquad (i=1,2,3) \;.
\label{tp3}
\end{eqnarray}
The last equation represents the redundancy in parameters explicitly identified
under the optimal parametrization (SVP) used. Hence, within the formulation,
the RPV parameters $B_i$'s and $\mu_i$'s are not exactly independent quantities.
Here, the $\tilde{m}^2_{\!{\scriptscriptstyle L}_{\!{\scriptscriptstyle 0}i}}$'s
play an interesting role very different from the R-parity conserving 
elements $\tilde{m}^2_{\!{\scriptscriptstyle L}_{ij}}$'s in the same matrix.
What values the 
$\tilde{m}^2_{\!{\scriptscriptstyle L}_{\!{\scriptscriptstyle 0}i}}$'s
might naturally take is quite a technical issue.
While the off-diagonal $\tilde{m}^2_{\!{\scriptscriptstyle L}_{ij}}$'s 
characterize LFV from soft SUSY breaking, the
$\tilde{m}^2_{\!{\scriptscriptstyle L}_{\!{\scriptscriptstyle 0}i}}$'s
could be so interpreted only when the SUSY breaking mechanism naturally
gives vanishing VEV's for the $\hat{L}_i$'s. Such a scenario is indeed
possible\cite{banks}. In a more generic situation, one can still think about
taking the low-energy Lagrangian and perform the necessary rotation to
recast the model into the SVP framework. Such a rotation may generate nonzero
$\tilde{m}^2_{\!{\scriptscriptstyle L}_{\!{\scriptscriptstyle 0}i}}$'s
to guarantee the satisfaction of Eq.(\ref{tp3}) in the new basis. Hence, when 
we switch off the $\tilde{m}^2_{\!{\scriptscriptstyle L}_{ij}}$'s to
single out the RPV contributions to LFV below, the 
$\tilde{m}^2_{\!{\scriptscriptstyle L}_{\!{\scriptscriptstyle 0}i}}$'s
will not be handled in the same fashion. We will discuss more the implications
of Eq.(\ref{tp3}) in relation to $\mu\to e\, \gamma$ below. We emphasize here
again that the above tadpole conditions are very important and should not 
be overlooked in any discussion of the phenomenology, particularly those 
related to the $B_i$'s.
 
\subsection{\protect\boldmath The
$\ell^{\!\!\mbox{ -}}_j  \to \ell^{\!\!\mbox{ -}}_i \, \gamma$
amplitude from (colorless) 1-loop diagrams}
Once we have the above matrices we can express the effective interactions involving
an external charged lepton with internal particles in terms of the mass eigenstates
and the elements of diagonalizing matrices.  The effective interaction for an 
external charged lepton, taken as an $l^{\!\!\mbox{ -}}_i$ 
or $l^{\scriptscriptstyle +}_i$  state for the $L$- or $R$-handed component here, 
with exact physical charged scalars and neutral fermions inside the loop is given by
\begin{equation}
\label{int1}
{\cal L} = {g_{\scriptscriptstyle 2}} \;
\overline{\Psi}({l_{i}}) 
\left[ {\cal N}^{\scriptscriptstyle L}_{inm} \,
 {1 - \gamma_{\scriptscriptstyle 5} \over 2} + 
{\cal N}^{\scriptscriptstyle R}_{inm} \, 
{1+ \gamma_{\scriptscriptstyle 5} \over 2} \right ] \, 
{\Psi}({\chi}^0_n) \, \phi_{m}^{\!\!\mbox{ -}} \;
+ \mbox{h.c.} \;,
\end{equation}
where ${1 \over 2}(1 \mp \gamma_{\scriptscriptstyle 5})$ are the $L$- and 
$R$-handed projections and 
\begin{eqnarray}
{\cal N}^{\scriptscriptstyle R}_{\scriptscriptstyle inm}
&=& \frac{1}{\sqrt{2}} \,       
[ \tan\!\theta_{\!\scriptscriptstyle W} 
\mbox{\boldmath $X$}_{\!\!1n}^{*} + \mbox{\boldmath $X$}_{\!\!2n}^{*} ] \,
{\cal D}^{l}_{(i+2)m} 
 - \frac{y_{\!\scriptscriptstyle e_i}}{g_{\scriptscriptstyle 2}} \,
\mbox{\boldmath $X$}_{\!\!4n}^{*} \, {\cal D}^{l}_{(i+5)m} 
- {\lambda_{kih}^{\!*} \over g_{\scriptscriptstyle 2} }\,\, 
\mbox{\boldmath $X$}_{\!(k+4)n}^{*} \, {\cal D}^{l}_{\!(h+5)m}  \;,
\label{Nrinm} \\
{\cal N}^{\scriptscriptstyle L}_{\scriptscriptstyle inm} 
&=& - \sqrt{2}\, \tan\!\theta_{\!\scriptscriptstyle W} 
\mbox{\boldmath $X$}_{\!\!1n} \, 
 {\cal D}^{l}_{(i+5)m} 
- \frac{y_{\!\scriptscriptstyle e_i}}{g_{\scriptscriptstyle 2}} \,
\mbox{\boldmath $X$}_{\!\!4n} {\cal D}^{l}_{(i+2)m} 
+ \frac{y_{\!\scriptscriptstyle e_i}}{g_{\scriptscriptstyle 2}} \,
 \mbox{\boldmath $X$}_{\!(i+4)n} \, {\cal D}^l_{2m}
- {\lambda_{khi} \over g_{\scriptscriptstyle 2} }\,\, 
\mbox{\boldmath $X$}_{\!(k+4)n} \, {\cal D}^{l}_{\!(h+2)m}  \; ,
\label{Nlinm}
\end{eqnarray}
with $n$ runs from 1 to 7 and $m$ from 1 to 8. We called this class of contributions
 neutralino-like, which obviously does include the ones with the physical 
neutralinos among the most important parts. The terms in 
${\cal N}^{\scriptscriptstyle L,R}_{\scriptscriptstyle inm}$ 
are easy to understand. For example, in
${\cal N}^{\scriptscriptstyle R}_{\scriptscriptstyle inm}$, the first two 
terms denote the interactions with $L$-handed sleptons and the bino and wino,
respectively.  The third term is the interaction with $R$-handed 
sleptons and the higgsino, while the fourth term describes the interaction 
with $R$-handed sleptons and the neutrino flavor states.  
Next, we come to the chargino-like contributions.
The corresponding effective interaction for the external charged lepton $l_{i}^\mp$
with exact physical neutral scalars and charged fermions inside the loop is given by
\begin{equation}
\label{int2}
{\cal L} = {g_{\scriptscriptstyle 2}} \;
\overline{\Psi}({l_{i}}) 
\left[ {\cal C}^{\scriptscriptstyle L}_{inm} \, 
{1 - \gamma_{\scriptscriptstyle 5} \over 2}  + 
{\cal C}^{\scriptscriptstyle R}_{inm} \,  
{1 + \gamma_{\scriptscriptstyle 5} \over 2}  \right] \, 
{\Psi}({\chi}^{\!\!\mbox{ -}}_n) \;
\phi_{m}^{\scriptscriptstyle 0} \; + \mbox{h.c.} \;,
\end{equation}
where
\begin{eqnarray}
 {\cal C}^{\scriptscriptstyle R}_{\scriptscriptstyle inm}
& =&  - \mbox{\boldmath $V$}_{\!\!1n} \,  {1 \over \sqrt{2}} \, 
[ {\cal D}^{s}_{\!(i+2)m} + i \, {\cal D}^{s}_{\!(i+7)m} ]
- \frac{y_{\!\scriptscriptstyle e_i}}{g_{\scriptscriptstyle 2}} \,
 \mbox{\boldmath $V$}_{\!\!(i+2)n}  \,  {1 \over \sqrt{2}} \, 
[ {\cal D}^{s*}_{\!2m} - i \, {\cal D}^{s*}_{\!7m} ] 
\nonumber \\ &&
- \frac{\lambda_{hik}^{\!*}}{g_{\scriptscriptstyle 2}} \,
\mbox{\boldmath $V$}_{\!\!(k+2)n} \,   {1 \over \sqrt{2}} \, 
[ {\cal D}^{s*}_{\!(h+2)m} - i \, {\cal D}^{s*}_{\!(h+7)m} ] \; ,
\label{Crinm} \\
 {\cal C}^{\scriptscriptstyle L}_{\scriptscriptstyle inm}
& = &   \frac{y_{\!\scriptscriptstyle e_i}}{g_{\scriptscriptstyle 2}} \,
  \mbox{\boldmath $U$}_{\!2n} \,  {1 \over \sqrt{2}} \, 
[ {\cal D}^{s}_{\!(i+2)m} + i \, {\cal D}^s_{\!(i+7)m} ]
- \frac{y_{\!\scriptscriptstyle e_i}}{g_{\scriptscriptstyle 2}} 
\mbox{\boldmath $U$}_{\!(i+2)n} \,  {1 \over \sqrt{2}} \, 
[ {\cal D}^s_{\!2m} + i \, {\cal D}^s_{\!7m} ]
\nonumber \\ &&
+ {\lambda_{khi} \over g_{\scriptscriptstyle 2} } \, 
\mbox{\boldmath $U$}_{\!(k+2)n} \,  {1 \over \sqrt{2}} \, 
 [ {\cal D}^{s}_{\!(h+2)m} + i \, {\cal D}^s_{\!(h+7)m} ] \; .
\label{Clinm}
\end{eqnarray}
Recall that ${\cal D}^{s}$ is actually real, though we are using ${\cal D}^{s*}$
notation as if it is not. This is just a convention for tracing the LFV 
structure of the various contributions in our analytical discussions below.
Here, in fact, the real difference between the ${\cal D}^{s*}$ and ${\cal D}^{s}$
terms is given explicitly by the different signs between the corresponding
scalar and pseudoscalar parts. Note that the ${y_{\!\scriptscriptstyle e_i}}$ terms
in the above expressions can be written together with the $\lambda$ terms 
using the $\lambda_{{\scriptscriptstyle \alpha \beta}k}$ notation and the
identification of ${y_{\!\scriptscriptstyle e_i}}$ 
as $\lambda_{{\scriptscriptstyle 0}ii}$. This common structure between
$\hat{L}_0$ and the $\hat{L}_i$'s is very useful in our discussions below.

In applying the above interactions to the process
$\ell^{\!\!\mbox{ -}}_j (p) \to \ell^{\!\!\mbox{ -}}_i \, \gamma (q)$, 
we can write the amplitude as
\begin{equation}
T = e \; \epsilon^{*\scriptscriptstyle \alpha} \, \bar{u}_i (p-q)
\left[
m_{\scriptscriptstyle \ell_{\!j}} \,
i \,\sigma_{\!\alpha \beta} \, q^\beta 
\left(
A_2^{\!\scriptscriptstyle L}  \, 
\frac{1-\gamma_{\scriptscriptstyle 5}}{2}
+ A_2^{\!\scriptscriptstyle R} \, 
\frac{1+\gamma_{\scriptscriptstyle 5}}{2}
\right )  \right ] u_j(p) \; ,
\end{equation}
where $\epsilon^*=\epsilon^*(q)$ is the polarization four vector of the
outgoing photon.
The decay rate is then simply given by
\begin{equation}
\Gamma({\ell}_{\!\scriptscriptstyle 2}^{\!\!\mbox{ -}} \to 
{\ell}_{\!\scriptscriptstyle 1}^{\!\!\mbox{ -}} \, \gamma)
= {\alpha_{\mbox{\tiny em}} \over 4 } \, m_{\!\mu}^{5} \,
\left( \,
|A_2^{\!\scriptscriptstyle L}|^2+|A_2^{\!\scriptscriptstyle R}|^2
\, \right) \; .
\end{equation}
It is straightforward to calculate the contributions from 1-loop diagrams with the 
effective interactions of Eqs. (\ref{int1}) and (\ref{int2}).  The result for 
 $A_2^{\!\scriptscriptstyle L}$ 
($A_2^{\!\scriptscriptstyle R}=A_2^{\!\scriptscriptstyle L}|_{L \leftrightarrow R}$) 
is given by
\begin{eqnarray}
A_2^{\!\scriptscriptstyle L}&=& \,
{\alpha_{\mbox{\tiny em}} \over 8 \pi \, 
\sin\!^2\theta_{\!\scriptscriptstyle W}} \;
\frac{1}{M_{\!\scriptscriptstyle \tilde{\ell}_{m}}^2}
 \left[
 {\cal N}_{\!\scriptscriptstyle inm}^{\scriptscriptstyle L} \,
 {\cal N}_{\!\scriptscriptstyle jnm}^{\scriptscriptstyle L^*} \,  
 F_2\!\!\left({{M}_{\!\scriptscriptstyle \chi^0_{n}}^2 \over M_{\!\scriptscriptstyle \tilde{\ell}_{m}}^2} \right)
+ {\cal N}_{\!\scriptscriptstyle inm}^{\scriptscriptstyle R} \,
 {\cal N}_{\!\scriptscriptstyle jnm}^{\scriptscriptstyle R^*} \,  
\; \frac{m_{\scriptscriptstyle \ell_{\!i}}}{m_{\scriptscriptstyle \ell_{\!j}}} \,
 F_2\!\!\left({{M}_{\!\scriptscriptstyle \chi^0_{n}}^2 \over M_{\!\scriptscriptstyle \tilde{\ell}_{m}}^2} \right)
+  {\cal N}_{\!\scriptscriptstyle inm}^{\scriptscriptstyle L} \,
{\cal N}_{\!\scriptscriptstyle jnm}^{\scriptscriptstyle R^*}  \, 
 \, {{M}_{\!\scriptscriptstyle \chi^0_{n}} 
 \over m_{\scriptscriptstyle \ell_{\!j}}} \,
F_3\!\!\left({{M}_{\!\scriptscriptstyle \chi^0_{n}}^2 \over M_{\!\scriptscriptstyle \tilde{\ell}_{m}}^2} \right)
\right] \nonumber \\
&-& {\alpha_{\mbox{\tiny em}} \over 8 \pi \,
 \sin\!^2\theta_{\!\scriptscriptstyle W}} \;
\frac{1}{M_{\!\scriptscriptstyle S_{m}}^2} \, 
\left[
 {\cal C}_{\!\scriptscriptstyle inm}^{\scriptscriptstyle L} \,
{\cal C}_{\!\scriptscriptstyle jnm}^{\scriptscriptstyle L^*} \, 
F_5\!\!\left({{M}_{\!\scriptscriptstyle \chi^{\mbox{-}}_{n}}^2 \over 
M_{\!\scriptscriptstyle S_{m}}^2} \right) 
+  {\cal C}_{\!\scriptscriptstyle inm}^{\scriptscriptstyle R} \,
{\cal C}_{\!\scriptscriptstyle jnm}^{\scriptscriptstyle R^*} \, 
\frac{m_{\scriptscriptstyle \ell_{\!i}}}{m_{\scriptscriptstyle \ell_{\!j}}} \,
F_5\!\!\left({{M}_{\!\scriptscriptstyle \chi^{\mbox{-}}_{n}}^2 \over 
M_{\!\scriptscriptstyle S_{m}}^2} \right) 
+{\cal C}_{\!\scriptscriptstyle inm}^{\scriptscriptstyle L}\,
 {\cal C}_{\!\scriptscriptstyle jnm}^{\scriptscriptstyle R^*} \,  
{{M}_{\!\scriptscriptstyle \chi^{\mbox{-}}_{n}} \over m_{\scriptscriptstyle \ell_{\!j}}} \, 
F_6\!\!\left({{M}_{\!\scriptscriptstyle \chi^{\mbox{-}}_{n}}^2 \over 
M_{\!\scriptscriptstyle S_{m}}^2} \right) 
 \right], \label{A2L}
\end{eqnarray}
where
\begin{eqnarray} 
F_2(x) &=& \frac{1}{6 \, (1-x)^4} \, (1-6 \, x+3 \, x^2+2 \, x^3-6 \, x^2 \, \ln x) \; ,
\nonumber \\
F_3(x) &=& \frac{1}{(1-x)^3} \, (1-x^2+2 \, x \,\ln x) \; ,
\nonumber \\
F_5(x) &=& 
\frac{1}{6 \,(1-x)^4} \, (2+3 \, x-6 \, x^2+x^3+6 \, x  \, \ln x) \;,
\nonumber \\
F_6(x) &=&
\frac{1}{(1-x)^3} \, (-3+4 \, x-x^2-2 \,  \ln x) \; ,
\nonumber 
\end{eqnarray}
with summations over all physical fermion and scalar mass eigenstates as 
represented by the $n$ and $m$ indices assumed. Apart from the model being different,
our background notation here, such as the loop integral functions, follows mostly
that of Ref.\cite{hmty}, to which readers are referred for a comparison. 

\section{Analysis of the various contributions}
Comparing ${\cal C}^{\scriptscriptstyle L,R}$ and 
${\cal N}^{\scriptscriptstyle L,R}$, we see that the two types of loop 
contributions as given in Eq.(\ref{A2L}) do have very similar structures.
The first type, corresponding to diagrams with a charged fermion
and neutral scalar in the loop, are, however, typically larger than 
their $SU(2)$ counterparts, {\it i.e.},
from diagrams with neutral fermions and charged scalars. Hence, we focus
our discussions on the ${\cal C}^{\scriptscriptstyle L,R}$ part, the
chargino-like loop. Before going into the analysis, it is instructive to 
introduce the lepton-flavor numbers $L_e$, $L_\mu$, and $L_\tau$ to the 
superfields as one does to their corresponding components in the SM. 
Some of the RPV parameters would then bear 
violations of the lepton-flavor numbers. It is obvious that in order to
have a contribution to $\mu^{\!\!\mbox{ -}} \to e^{\!\!\mbox{ -}}\,\gamma$,
a term must reduce $L_\mu$ and increase $L_e$ by exactly
one unit while leaving $L_\tau$ unchanged. This simple but useful rule 
serves as a countercheck of individual contributions discussed below.

The following discussions are concerned with sophisticated
analytical details on the more interesting individual pieces of contributions.
The aim is to understand the exact role played by each of the RPV parameters
in the process, and the strength of each contribution. Reading the section
would take a bit of effort, which is not necessary for a general reader
interested mainly in knowing the basic features of the results. While we
will refer to some analytical results presented here in the discussion 
of the numerical results in the next section, for comparison and confirmation,
a general reader may simply skip this section.

\subsection{Chirality flip inside the loop}
Take $A^{\!\scriptscriptstyle L}_2$, for example, among the three terms in 
Eq.(\ref{A2L}), the one with ${\cal C}^{\scriptscriptstyle L}_{1nm}\, 
{\cal C}^{\scriptscriptstyle R^*}_{2nm}$ 
corresponds to the diagrams with the chirality flip on the internal
fermion line. Hence, the diagrams with the two chargino states ({\it i.e.}, $n=1,2$)
would potentially dominate over the other diagrams with external
chirality-flipping mass insertion ($m_\mu$ or $m_e$). 

We first look at the  contributions with a 
($\mu^*\lambda$) structure. Taking only one $\lambda$-coupling vertex, 
one can then take the gauge coupling term (first term) from
${\cal C}^{\scriptscriptstyle R^*}_{2nm}$. We then have the real scalar part 
of the contribution proportional to
\beq \label{1st}
\sum_{n=1}^5 \sum_{m=1}^{5} \, \mbox{\boldmath $V$}^{\!*}_{\!\!1n}\,
{M}_{\!\scriptscriptstyle \chi^{\mbox{-}}_n} 
 \mbox{\boldmath $U$}_{\!(k+2)n} \; 
F_6\!\!\left({{M}_{\!\scriptscriptstyle \chi^{\mbox{-}}_{n}}^2 \over 
M_{\!\scriptscriptstyle S_{m}}^2} \right)  \;
{\cal D}^{s*}_{\!4m} \, {\cal D}^{s}_{\!(h+2)m} \;
\frac{\lambda_{kh\scriptscriptstyle 1}}{g_{\scriptscriptstyle 2}}\; \; .
\eeq
Here we have dropped the pseudoscalar part just for simplicity in this 
discussion. This is exactly valid when there is no mixings between scalars and 
pseudoscalars. However, we find it more transparent in illustrating the basic
features. If the loop function $F_6$ could be factored out from the double 
summation, we would have a $\mbox{\boldmath $V$}^{\!*}_{\!\!1n}\,
{M}_{\!\scriptscriptstyle \chi^{\mbox{-}}_n} 
 \mbox{\boldmath $U$}_{\!(k+2)n}$ 
summation over fermions and a 
${\cal D}^{s*}_{\!4m} \, {\cal D}^{s}_{\!(h+2)m}$ summation over (real) 
scalars. However, the fermionic sum gives exactly the
${l}_k^{\!\!\mbox{ -}}$--$\tilde{W}^{\scriptscriptstyle +}$ mass term, which is 
zero. This is well illustrated by the loop diagram with electroweak state notation, 
as given in Fig.~1. Hence, one expects a GIM-like cancellation mechanism, here  
violated only to the extent that $F_6$ is not universal. The violation is, however,
quite substantial, as illustrated in the similar case of quark electric dipole
moments\cite{as6} and by our exact numerical calculation here. The case of 
degenerate charginos is an obvious exception, which, however, needs too large a 
complex phase for $\mu_{\scriptscriptstyle 0}$ to be phenomenologically viable.
A similar situation goes with the scalar sum --- 
${\cal D}^{s*}_{\!4m} \, {\cal D}^{s}_{\!(h+2)m} = \delta_{h2}$ by unitarity.
Taking $h=2$ in the above expression (\ref{1st}), we have the two dominating 
chargino contributions, the $n=1$ and $2$ parts, given approximately by
\beq \label{1st+}
\mbox{\boldmath $V$}^{\!*}_{\!\!1n}\,
{\mu_k^*} \, R_{\!\scriptscriptstyle R_{2n}}  \,
\frac{\lambda_{k\scriptscriptstyle 21}}{g_{\scriptscriptstyle 2}} \;
F_6\!\!\left({{M}_{\!\scriptscriptstyle \chi^{\mbox{-}}_{n}}^2 \over 
M_{\!\scriptscriptstyle S_{m}}^2} \right)  \;;
\eeq
where $R_{\!\scriptscriptstyle R}$ is a $2\times 2$ matrix with order one 
matrix elements and is the $R$-handed transformation needed to diagonalize the 
$2\times 2$ MSSM chargino block (see Appendix A of Ref.\cite{as6} for details.)
The expected combination $\mu_k^* \,\lambda_{k\scriptscriptstyle 21}$ comes
up, with $k=1$ and $3$ admissible. There is only a partial cancellation between the 
two parts in general. The same situation goes for the
${\cal C}^{\scriptscriptstyle R}_{1nm}\, {\cal C}^{\scriptscriptstyle L^*}_{2nm}$ 
part of  $A^{\!\scriptscriptstyle R}_2$, with the combination 
$\mu_k \,\lambda_{k\scriptscriptstyle 12}^*$  (~$k=2$ and $3$ admissible here~)
instead. An interesting point to note is that the above expression
shows no obvious dependence on $\tan\!\beta$, a result confirmed by our 
exact numerical calculation. This is an important issue that we will get back 
to below.
 
Next, we consider the above expression (\ref{1st})
with $h\ne 2$. Let us try to estimate the 
individual terms in the ${\cal D}^{s*}_{\!4m} \, {\cal D}^{s}_{\!(h+2)m}$ summation.
From the structure of the neutral scalar mass matrix, we see that 
${\cal M}^2_{\!\scriptscriptstyle SS}$ 
in particular, is very likely to be predominantly diagonal. Off-diagonal 
elements in the  ${\cal D}^{s}$ matrix then 
measure the small mixings. The largest terms among the different $m$'s are then
given by $m=4$ or $m=h+2$ with a suppression factor of magnitude
${\cal D}^{s}_{\!(h+2)4}={\cal D}^{s*}_{\!4(h+2)}\sim 
\frac{ \mu_{\scriptscriptstyle 2}\, \mu_h^* }{\tilde{m}^2_{L_{hh}}
- \tilde{m}^2_{L_{22}}}$. This result is an example of the $L$-slepton
flavor mixing from the coupling of the form $\mu_i^* \,  \mu_j$ mentioned in
the Introduction --- something that is also useful for our discussion
below. The flavor mixing is explicitly given in Eq.(\ref{Mp}), in  
the discussion of the scalar masses, where we explain the notation 
${\cal D}^{s*}$ despite its being real. As the RPV mixings are the same in the
scalar and pseudoscalar parts (${\cal M}^2_{\!\scriptscriptstyle SS}$ and
${\cal M}^2_{\!\scriptscriptstyle PP}$), one could think in terms of complex 
scalars here, in which case the complex ${\cal D}^{s}$ notation would be 
exactly valid. With a first-order difference between 
$\tilde{m}^2_{\!\scriptscriptstyle L_{hh}}$ and 
$\tilde{m}^2_{\!\scriptscriptstyle L_{22}}$, 
the nonuniversal $F_6$ function still induce large
violation in the unitarity cancellation, essentially between the $m=4$ and $m=h+2$
parts. For the case at hand, the contribution goes with four, instead of
two, RPV parameters (the two combinations 
${ \mu_{\scriptscriptstyle 2}\, \mu_h^*}$ and 
$\mu_k^* \,\lambda_{kh\scriptscriptstyle 1}$); hence, it is very likely to be 
subdominating.

There are analogous neutralino-like contributions. However, in the latter case, 
there is also an extra pure gaugino loop, from the first terms in Eqs.(\ref{Nrinm}) 
and (\ref{Nlinm}) with no parallel chargino-like counterparts. 
This contribution involves ${\cal D}^{l*}_{\!4m} \, {\cal D}^{l}_{\!6m}$ or
${\cal D}^{l*}_{\!7m} \, {\cal D}^{l}_{\!3m}$,
hence a $LR$-slepton mixing, with RPV contribution given by 
$\mu_k^* \,\lambda_{k\scriptscriptstyle 21}$ or
$\mu_k\,\lambda_{k\scriptscriptstyle 12}^* $  
(for $A^{\!\scriptscriptstyle L}_2$ or $A^{\!\scriptscriptstyle R}_2$,
respectively). This is the most intuitive contribution involving the combination 
of parameters that we mentioned in the introduction. Numerically, this contribution 
is, typically, only similar to other neutralino-like terms with one gauge 
coupling vertex,  and hence smaller than the dominating chargino-like term 
discussed above. The reason here is that the larger gauge coupling effect is 
offset by the suppression factor coming from the $LR$-mixing. However, if 
one pushes for a large $|\mu_{\scriptscriptstyle 0}|$ value while keeping the
bino mass $M_{\scriptscriptstyle 1}$ small, the other contributions, including
the whole chargino-like loop contribution could become suppressed, thus
leaving the pure gauge loop to be the dominating one.

Replacing the $\lambda$ coupling in the above contributions
[{\it cf.} expression (\ref{1st})] with a ``leptonic Yukawa coupling" 
(recall  ${y_{\!\scriptscriptstyle e_i}} \equiv \lambda_{{\scriptscriptstyle 0}ii}$),
we have a
\beq \label{2murl1}
\sum_{n=1}^5 \sum_{m=1}^{5} \, \mbox{\boldmath $V$}^{\!*}_{\!\!1n}\,
{M}_{\!\scriptscriptstyle \chi^{\mbox{-}}_n} 
 \mbox{\boldmath $U$}_{\!2n} \; 
F_6\!\!\left({{M}_{\!\scriptscriptstyle \chi^{\mbox{-}}_{n}}^2 \over 
M_{\!\scriptscriptstyle S_{m}}^2} \right)  \;
{\cal D}^{s*}_{\!4m} \, {\cal D}^{s}_{\!3m} \; 
\frac{y_{\!\scriptscriptstyle e_1}}{g_{\scriptscriptstyle 2}}\; 
\eeq
part. Here, the fermionic sum has obviously a dominating chargino contribution 
without the need for RPV parameters. However, we have no choice but the fixed 
scalar mixings given by ${\cal D}^{s*}_{\!4m} \, {\cal D}^{s}_{\!3m}$. 
As discussed above, the latter involves 
${ \mu_{\scriptscriptstyle 1}^* \, \mu_{\scriptscriptstyle 2}}$. With the small
${y_{\!\scriptscriptstyle e_1}}$ coupling, 
it gives a strong suppression. Of course the ``electron Yukawa"
${y_{\!\scriptscriptstyle e_1}}$ above can be exchanged for the larger
``muon Yukawa" ${y_{\!\scriptscriptstyle e_2}}$ in the corresponding piece 
in  $A^{\!\scriptscriptstyle R}_2$ instead. The latter give the same 
${\cal D}^{s*}_{\!4m} \, {\cal D}^{s}_{\!3m}$ scalar mixing factor.
This contribution is depicted in Fig.~2. 

From Eqs.(\ref{Crinm}) and (\ref{Clinm}), there is one more similar 
but independent term. This has expression (\ref{2murl1}) modified with a 
simple switching of the explicit $2$ and $3$ indices, and a sign flip.
Here we write down explicitly the corresponding
term from $A^{\!\scriptscriptstyle R}_2$ instead :
\beq \label{2murl2}
\sum_{n=1}^5 \sum_{m=1}^{5} - \mbox{\boldmath $V$}_{\!\!1n}\,
{M}_{\!\scriptscriptstyle \chi^{\mbox{-}}_n} 
 \mbox{\boldmath $U$}^{\!*}_{\!4n} \; 
F_6\!\!\left({{M}_{\!\scriptscriptstyle \chi^{\mbox{-}}_{n}}^2 \over 
M_{\!\scriptscriptstyle S_{m}}^2} \right)  \;
{\cal D}^{s}_{\!3m} \, {\cal D}^{s*}_{\!2m} \; 
\frac{y_{\!\scriptscriptstyle e_2}}{g_{\scriptscriptstyle 2}}\; .
\eeq
From the above discussion, it is easy to see that the dominating part with
the charginos ($n=1$ and $2$) gives a $\mu_{\scriptscriptstyle 2}$ dependence through
$\mbox{\boldmath $U$}^*_{\!4n}$. Naively, we expect a
$\mu_{\scriptscriptstyle 1}^*$ from the scalar mixing part. The mixing of
the corresponding complex scalars, however, involves a 32- entry of the 
${\cal M}_{\!\scriptscriptstyle {\phi}{\phi}^\dag}^2$ matrix. From the result
given in the Sec.~IIA, we see that the mixing does have a
$\mu_{\scriptscriptstyle 1}^*$. But it comes in the combination
$\tilde{m}^2_{\!{\scriptscriptstyle L}_{\!{\scriptscriptstyle 10}} }
+ \mu_{\scriptscriptstyle 1}^{*}\,\mu_{\scriptscriptstyle 0}$  
The latter is, by the tadpole equation (\ref{tp3}), $B_1^* \, \tan\!\beta$.
Detailed analysis of the full scalar mass matrix, with the first-order mixings
among the Higgs states, does not change the $B_1^* \, \tan\!\beta$
dependence of the result. Hence, the LFV structure here is given by
the RPV parameter combination $B_1^*\,\mu_{\scriptscriptstyle 2}$, which
is related to the  $\mu_{\scriptscriptstyle 1}^{*}\,\mu_{\scriptscriptstyle 2}$
combination through the tadpole equation. Given in terms of $B_1^*$,
the contribution has an explicit $\tan\!\beta$ dependence. In general, one
expects this contribution to be of comparable magnitude to the previous one.

The last contribution we want to discuss here is the one with two
$\lambda$ couplings. Extracting the part similar to expression (\ref{1st}),
we have
\beq \label{maj}
\sum_{m}^{\prime}
\sum_{n=1}^{5} \, \mbox{\boldmath $V$}^{\!*}_{\!\!(k'+2)n}\,
{M}_{\!\scriptscriptstyle \chi^{\mbox{-}}_n} 
 \mbox{\boldmath $U$}_{\!(k+2)n} \; 
F_6\!\!\left({{M}_{\!\scriptscriptstyle \chi^{\mbox{-}}_{n}}^2 \over 
M_{\!\scriptscriptstyle S_{m}}^2} \right)  
[{\cal D}^{s}_{\!(h'+2)m} + i {\cal D}^{s}_{\!(h'+7)m} ]
\, [{\cal D}^{s}_{\!(h+2)m} + i {\cal D}^{s}_{\!(h+7)m} ] \;  
\frac{\lambda_{h'{\scriptscriptstyle \!2}k'}}{g_{\scriptscriptstyle 2}}\;
\frac{\lambda_{kh\scriptscriptstyle 1} }{g_{\scriptscriptstyle 2}}\; .
\eeq
Note that we have written the expression very differently here. Both the scalar
and pseudoscalar parts are explicitly shown. The sum over $m$,
scalar mass eigenstates, goes over all nine physical states (the $\sum_{m}^{\prime}$
notation means the unphysical Goldstone mode is omitted). The interesting point 
here is that this contribution involves a nontrivial interplay between two parts. 
A careful reading of the $\lambda_{h'{\scriptscriptstyle \!2}k'}^*$ term in 
Eq.(\ref{Crinm}) 
would appreciate that the scalar mixing matrix comes in as ${\cal D}^{s*}$ instead, 
or, equivalently, the scalar and pseudoscalar parts come in with the ``wrong" sign.
At the limit of degenerate but unmixed scalars and pseudoscalars, the ``right" signs 
(or with ${\cal D}^{s}$, as in the case of the first gauge coupling term) would
give identical contributions from the two parts; while the ``wrong" signs would 
give an exact cancellation instead. To put it in another way, a Majorana-like
scalar mass insertion along the scalar line is needed to have the contribution
nonzero, as depicted in Fig.~3. There have been some discussions on the 
Majorana-like scalar mass terms for the ``sneutrinos"
(or $\tilde{l}_i^{\scriptscriptstyle 0}$ states to be exact)  as complex scalars,
and some of the resulting phenomenological implications are studied
in recent literature\cite{Bs,Bnu}. Such mass terms appear under the SVP 
through LFV from the soft $B_i$ parameters.  An illustration of the Majorana-like 
scalar mass terms is shown explicitly in Fig.~3b. It can be easily seen that
the contribution under discussion here actually involves a minimum of four RPV 
parameters, $B_{h'}^* \, B_h^*$ and 
${\lambda_{h'{\scriptscriptstyle \!2}k'}}\,{\lambda_{kh\scriptscriptstyle 1} }$,
and it is achieved by taking $k=k'$. An interesting point,
at least from the theoretical point of view, is that
the contribution could have a different flavor structure from the more
familiar two-$\lambda$ loop diagrams with the chirality flip on the
external muon line\cite{CH}, and a $m_{\tau}$ instead of 
$m_{\mu}$ dependence too. For instance, we could have a new contribution from the 
$\lambda_{{\scriptscriptstyle 1\!23}} \, \lambda_{{\scriptscriptstyle 31\!1}}$
combination.

Finally, we want to remark that the remaining contributions with internal 
chirality flip can be analyzed again by using the extended flavor structure
with the ``leptonic Yukawa coupling"  ${y_{\!\scriptscriptstyle e_i}}$
identified as $\lambda_{{\scriptscriptstyle 0}ii}$, as discussed above.
In fact, Fig.~3a is given with the generic $\lambda_{{\scriptscriptstyle \za \zb}k}$
notation and hence applicable to this latter case. One can obtain contributions 
depending only on a $B_h^*$ and $\lambda_{{h\scriptscriptstyle 21}}$
combination, for example. Explicitly, we have expression (\ref{maj}) modified
to 
\beq \label{Blam}
\sum_{m}^{\prime}
\sum_{n=1}^{5} \, \mbox{\boldmath $V$}^{\!*}_{\!\!4n}\,
{M}_{\!\scriptscriptstyle \chi^{\mbox{-}}_n} 
 \mbox{\boldmath $U$}_{\!4n} \; 
F_6\!\!\left({{M}_{\!\scriptscriptstyle \chi^{\mbox{-}}_{n}}^2 \over 
M_{\!\scriptscriptstyle S_{m}}^2} \right)  
[{\cal D}^{s}_{\!2m} + i {\cal D}^{s}_{\!7m} ]
\, [{\cal D}^{s}_{\!(h+2)m} + i {\cal D}^{s}_{\!(h+7)m} ] \;  
\frac{y_{\!\scriptscriptstyle e_2}}{g_{\scriptscriptstyle 2}}\;
\frac{-\lambda_{h\scriptscriptstyle 21} }{g_{\scriptscriptstyle 2}}\; ,
\eeq
with which the fermionic sum suggests a major contribution from $n=4$, {\it i.e.},
the muon itself with the $m_\mu$ dependence, and the scalar sum gives the
dominating contribution proportional to $B_h^*$. However, this contribution then
has two factors of ``muon Yukawa" (${y_{\!\scriptscriptstyle e_2}}$) suppression.
Hence, it is not a very important contribution.
Note that the Majorana-like scalar mass required can be interpreted as in the 
same form as given by Fig.~3b, with a $B_0^*$ and a $B_h^*$, except that the $B_0^*$
insertion, in general, has no suppression; the $B_0^*$ is actually not 
necessary to complete the corresponding diagram in Fig.~3a in this case though,
due to the existence of the $\hat{L}_0$ (or, equivalently 
$\tilde{l}_{\scriptscriptstyle 0}^{\scriptscriptstyle 0}\equiv 
h_{\scriptscriptstyle d}$) VEV. 

\subsection{Chirality flip on the external muon line}
Contributions from the terms with the chirality flip on the external muon line
is expected to be suppressed by a ${y_{\!\scriptscriptstyle e_2}}$ ($\sim 10^{-3}$)
factor. However, we have seen that apart from the $\mu^* \, \lambda$-type
combination, other combinations of RPV parameters that come into the diagrams 
with an internal chirality flip do have extra suppressions. Hence, it is still of
interest to see if there are important terms with an external chirality flip
that are comparable to those discussed in the previous section. Therefore,
we mainly look for terms depending on only two RPV parameters without further 
extra suppressions, apart from the one dictated by the chirality flip. 

For the chargino-like diagrams, for instance, our candidates here are from the
${\cal C}_{\!\scriptscriptstyle 1nm}^{\scriptscriptstyle L} \,
{\cal C}_{\!\scriptscriptstyle 2nm}^{\scriptscriptstyle L^*}$
part of $A_2^{\!\scriptscriptstyle L}$ and the 
${\cal C}_{\!\scriptscriptstyle 1nm}^{\scriptscriptstyle R} \,
{\cal C}_{\!\scriptscriptstyle 2nm}^{\scriptscriptstyle R^*}$
part of $A_2^{\!\scriptscriptstyle R}$. The latter looks more interesting.
First of all, there is a pure gauge loop, as shown in Fig.~4a. Compared with 
the above, we have the analogous expression
\beq \label{ext12}
m_{\scriptscriptstyle \mu}
\sum_{n=1}^5 \sum_{m=1}^{5} - \mbox{\boldmath $V$}^{\!*}_{\!\!1n}\,
 \mbox{\boldmath $V$}_{\!\!1n} \; 
F_5\!\!\left({{M}_{\!\scriptscriptstyle \chi^{\mbox{-}}_{n}}^2 \over 
M_{\!\scriptscriptstyle S_{m}}^2} \right)  \;
{\cal D}^{s*}_{\!4m} \, {\cal D}^{s}_{\!3m} \; .
\eeq
The scalar mixing part gives the 
$\mu_{\scriptscriptstyle 1}^*\,\mu_{\scriptscriptstyle 2}$ LFV structure. This 
looks comparable with expression (\ref{2murl1}). 

The only other term from 
${\cal C}_{\!\scriptscriptstyle 1nm}^{\scriptscriptstyle R} \,
{\cal C}_{\!\scriptscriptstyle 2nm}^{\scriptscriptstyle R^*}$
without further ${y_{\!\scriptscriptstyle e_i}}$ suppression is the usual
$2$-$\lambda$ term discussed much earlier in the literature, together with similar 
(colored) $2$-$\lambda^{\!\prime}$ diagrams\cite{CH}. We have an expression given by 
\beq \label{2lam}
m_{\scriptscriptstyle \mu}
\sum_{n=1}^5 \sum_{m=1}^{5} - \mbox{\boldmath $V$}^{\!*}_{\!\!(k'+2)n}\,
 \mbox{\boldmath $V$}_{\!\!(k+2)n} \; 
F_5\!\!\left({{M}_{\!\scriptscriptstyle \chi^{\mbox{-}}_{n}}^2 \over 
M_{\!\scriptscriptstyle S_{m}}^2} \right)  \;
{\cal D}^{s}_{\!(h'+2)m} \, {\cal D}^{s*}_{\!(h+2)m} \;
\frac{\lambda_{h'{\scriptscriptstyle \!2}k'}}{g_{\scriptscriptstyle 2}}\;
\frac{\lambda_{h{\scriptscriptstyle 1}k}^{\!*} }{g_{\scriptscriptstyle 2}}\; ,
\eeq
requiring further $h=h'$ and $k=k'$. Note that all off-diagonal matrix elements of
the form $\mbox{\boldmath $V$}_{\!\!(k+2)n}$ are very small, those RPV ones 
($n=1$ or 2) in particular further contain  a
${y_{\!\scriptscriptstyle e_i}}$ suppression (see Appendix A of Ref.\cite{as6}). 
Finally, with the ${\cal C}_{\!\scriptscriptstyle 1nm}^{\scriptscriptstyle L} \,
{\cal C}_{\!\scriptscriptstyle 2nm}^{\scriptscriptstyle L^*}$
part of $A_2^{\!\scriptscriptstyle L}$, we are obviously left only with the
other $2$-$\lambda$ terms\cite{CH}.

The neutralino-like loop contributions are mostly straight forward analogs, 
except for the contributions with only one gauge coupling.
Unlike the chargino-like case discussed above, there may be no extra 
${y_{\!\scriptscriptstyle e_i}}$ suppression, from a $\mbox{\boldmath $V$}$
element or otherwise. However, an extra $LR$-mixing is involved, giving another
suppression, as illustrated in Fig.~4b.

The only relatively important type of contributions involving 
the combination such as $B_h^* \, \lambda_{{h\scriptscriptstyle 21}}$
has two factors of ${y_{\!\scriptscriptstyle e_2}}$ suppression.
If one is interesting in the dominating contribution of this type, an external 
chirality flip term with only one extra ${y_{\!\scriptscriptstyle e_2}}$ suppression
may have to be considered. In fact, such a contribution does exist. The
 ${\cal C}_{\!\scriptscriptstyle 1nm}^{\scriptscriptstyle L} \,
{\cal C}_{\!\scriptscriptstyle 2nm}^{\scriptscriptstyle L^*}$
part of $A_2^{\!\scriptscriptstyle L}$ has a piece of the form 
\beq \label{Blam2}
m_{\scriptscriptstyle \mu}
\sum_{n=1}^5\sum_{m=1}^{5} - \mbox{\boldmath $U$}^{\!*}_{\!4n}\,
 \mbox{\boldmath $U$}_{\!4n} \; 
F_5\!\!\left({{M}_{\!\scriptscriptstyle \chi^{\mbox{-}}_{n}}^2 \over 
M_{\!\scriptscriptstyle S_{m}}^2} \right)  \;
{\cal D}^{s}_{\!(h+2)m} \, {\cal D}^{s*}_{\!2m} \;
\frac{- \lambda_{h\scriptscriptstyle 2\!1}}{g_{\scriptscriptstyle 2}}\;
\frac{- y_{\!\scriptscriptstyle e_2}}{g_{\scriptscriptstyle 2}}\; ,
\eeq
where we have $h=3$ or $1$. This is expected to be of similar strength to 
the one in expression (\ref{Blam}) discussed above. Recall that the scalar
mixings give a $B_h^*\tan\!\zb$ dependence.

\section{Exact Numerical calculations and results}
In this section, we present the results we obtained by a careful 
numerical implementation of our $\mu\to e \,\gamma$ formulas, with explicit 
numerical diagonalization of all the mass matrices involved. We isolate various 
major contributions by singling out each of the corresponding RPV parameter 
combinations as the only nonvanishing ones.  The soft
SUSY breaking contributions to R-parity conserving slepton mixings are set to
zero (~{\it i.e.}, $\tilde{m}^2_{\!\scriptscriptstyle L}$, 
$\tilde{m}^2_{\!\scriptscriptstyle E}$, and $A^{\!\scriptscriptstyle E}$ are 
set to be diagonal~). Though we have used only real numbers for all input parameters
in the numerical results presented, our discussions apply to the general case of
complex parameters, as does the analysis given above. A basic set of typical values 
chosen for the input parameters are given in Table~\ref{table1}. We used this set of
inputs unless otherwise specified in the results below.  At the end, we also show
the effects of varying these input parameters.

At the beginning of the previous section, we introduced a useful rule
in terms of lepton-flavor number violation counts; {\it i.e.},
an admissible contribution has to include RPV parameters such that it
reduces $L_\mu$ and increases $L_e$ by exactly one unit while leaving $L_\tau$ 
untouched. Moreover, $\mu\to e \,\gamma$ is an R-parity even process.
The rule alone can be used to identify a few interesting cases
of two RPV parameter combinations. We first concentrate on the superpotential
parameters. Using the bilinear parameters, we have only the
$\mu_{\scriptscriptstyle 1}^*\, \mu_{\scriptscriptstyle 2}$ combination. 
With the trilinear parameters, there are various $2$-$\lambda$ and 
$2$-$\lambda^{\!\prime}$ combinations. Such contributions to
$\mu\to e\, \gamma$ are well studied\cite{CH}. It is easy to see that
as the $\lambda^{\!\prime}$ couplings involve quarks and squarks while the others
do not, there is no combination of a single $\lambda^{\!\prime}$ with a
coupling of the other types contributing to 1-loop $\mu\to e\, \gamma$. We skip
the quark-squark loop contributions here. The  $2$-$\lambda$ loops, however, is
an integral part of our (colorless loop) formulas. We will discuss such
contributions also for completeness and for easy comparison with previous works.
Most interestingly, however, are the combinations involving RPV parameters
of both the bilinear and trilinear types. They are
${\mu_k^*}\,{\lambda_{k\scriptscriptstyle 21}}$ with $k$ being $3$ or $1$,
and ${\mu_k}\,{\lambda_{k\scriptscriptstyle 12}^*}$ with $k$ being $3$ or $2$.

Next, we add into consideration the RPV parameters from soft SUSY breaking. 
The trilinear $A^\lambda$ terms have no role to play here, because under
the SVP such terms give only three-scalar interaction vertices, which obviously 
cannot be incorporated into any 1-loop $\mu\to e\, \gamma$ diagram. 
This leaves the $B_i$'s and 
$\tilde{m}^2_{\!{\scriptscriptstyle L}_{\!{\scriptscriptstyle 0}i} }$'s. 
However, as commented above in the last part of Sec.~IIA, we have the important
tadpole conditions in Eq.(\ref{tp3}) relating these two sets of parameters to the
${\mu_i}$'s. In the literature, ${\mu_i}$'s and $B_i$'s are usually discussed
as independent parameters while the
$\tilde{m}^2_{\!{\scriptscriptstyle L}_{\!{\scriptscriptstyle 0}i} }$'s
and the tadpole conditions are often overlooked. Recall that imposing the correct 
tadpole equations is crucial in getting the correct physical scalars.
If one is in favor of setting 
$\tilde{m}^2_{\!{\scriptscriptstyle L}_{\!{\scriptscriptstyle 0}i} }$'s
to zero, a $B_i$ value is then fixed by the corresponding ${\mu_i}$, and vice versa.
However, in order to highlight some of the analytical features discussed and 
to enable an easy comparison with previous 
studies of other phenomenological implications of the
$B_i$ in the literature, we will first discuss the ${\mu_i}$'s and $B_i$'s
separately as if they are independent. We will first single out a contribution
involving a ${\mu_i}$ by setting the  $B_i$ to zero and tuning
$\tilde{m}^2_{\!{\scriptscriptstyle L}_{\!{\scriptscriptstyle 0}i} }$ to satisfy
the tadpole condition before solving for the spectra of the physical scalars.
Likewise, we will tune a ${\mu_i}$ to zero to single out a $B_i$ effect. 
An interested reader can put the two otherwise related contributions,
for example a ${\mu_{\scriptscriptstyle 3}^*}\,{\lambda_{\scriptscriptstyle 321}}$
and a ${B_{3}^*}\,{\lambda_{\scriptscriptstyle 321}}$ contribution, together
by imposing the tadpole equation and picking whatever 
$\tilde{m}^2_{\!{\scriptscriptstyle L}_{\!{\scriptscriptstyle 0}i}}$ 
value deemed appropriate. We will also show a case of 
$\tilde{m}^2_{\!{\scriptscriptstyle L}_{\!{\scriptscriptstyle 0}i} }=0$
for the  ${\mu_{\scriptscriptstyle 1}^*}\, {\mu_{\scriptscriptstyle 2}}$
contribution at the end. Please be reminded that in the tadpole equation,
$B_i$ goes with a $\tan\!\zb$ factor. For the same ${\mu_i}$, $B_i$ will then be
suppressed by $\tan\!\zb$.

It is easy to see that under the strategy discussed,
the additional RPV parameter combinations of interest are
given by $B_1^*\, \mu_{\scriptscriptstyle 2}$,
$\mu_{\scriptscriptstyle 1}^*\, B_2$, and  $B_1^*\, B_2$, together with
${B_{k}^*}\,{\lambda_{k\scriptscriptstyle 21}}$ and
${B_{k}}\,{\lambda_{k\scriptscriptstyle 12}^*}$. These complete our list.
Now, we go into each of these combinations of RPV parameters.
Readers interested in more analytical details are urged to compare our discussions 
below with those presented in the previous section, the cross references to which 
are given inside square brackets ([~]'s).

\subsection{\boldmath\protect The ${\mu_k^*}\,{\lambda_{k\scriptscriptstyle 21}}$ 
or  ${\mu_k}\,{\lambda_{k\scriptscriptstyle 12}^*}$ contributions.}
This is the most interesting case, because it involves both the bilinear and
trilinear RPV couplings and also it is likely to give a much larger branching ratio.
Later, we will see that the bounds obtained on such parameter combinations
are comparable to what one could obtain by imposing a sub-eV bound on all
neutrino mass contributions (see, for example, Refs\cite{as1,as5,ok}). 
Without loss of generality, we 
take the ${\mu_{\scriptscriptstyle 3}^*}\,{\lambda_{\scriptscriptstyle 321}}$ 
combination for illustration. The dominant contribution
comes from the last term of Eq.(\ref{A2L}) as shown in Fig.~1 [also discussed in 
expression (\ref{1st})]. This is confirmed by our exact numerical calculation.  
Another interesting contribution comes from the pure neutral gaugino with 
${\mu_{\scriptscriptstyle 3}^*}\,{\lambda_{\scriptscriptstyle 321}}$ 
coming in through the $LR$-slepton mixing. There are also the neutralino 
contributions without $LR$-slepton mixing, the exact analog of the chargino ones. 
In a generic region of the parameter space, the chargino-like loop result
typically dominates over the neutralino-like loop result.
We plot contours of the resulting branching ratio as a function of 
(real) ${\mu_{\scriptscriptstyle 3}}$ and $\lambda_{\scriptscriptstyle 321}$
in Fig.~\ref{fig5}. The present experimental limit is also shown and the allowed 
region at 90\% C.L.  is shaded. The contours for the
other three combinations of RPV parameters, each taken alone, are essentially
the same. Recall that the corresponding dominating contribution for the two 
${\mu_k}\,{\lambda_{k\scriptscriptstyle 12}^*}$ combinations comes in via the
${\cal C}^{\scriptscriptstyle R}_{1nm}\, {\cal C}^{\scriptscriptstyle L^*}_{2nm}$ 
part of $A^{\!\scriptscriptstyle R}_2$ instead.    The 90\% C.L. upper limit on 
$|{\mu_k^*}\,{\lambda_{k\scriptscriptstyle 21}}|$ 
or  $|{\mu_k}\,{\lambda_{k\scriptscriptstyle 12}^*}|$
(normalized by $|\mu_{\scriptscriptstyle 0}|= 100\,\mbox{GeV}$) is
given by
\begin{equation}
\frac{|{\mu_{\scriptscriptstyle 3}^*}\,{\lambda_{\scriptscriptstyle 321}}|}
{|\mu_{\scriptscriptstyle 0}|}\;, \;\;\;
\frac{|{\mu_{\scriptscriptstyle 1}^*}\,{\lambda_{\scriptscriptstyle 121}}|}
{|\mu_{\scriptscriptstyle 0}|}\;, \;\;\;
\frac{|{\mu_{\scriptscriptstyle 3}}\,{\lambda_{\scriptscriptstyle 312}^*}|}
{|\mu_{\scriptscriptstyle 0}|}\;, \;\;\; 
\frac{|{\mu_{\scriptscriptstyle 2}}\,{\lambda_{\scriptscriptstyle 212}^*}|}
{|\mu_{\scriptscriptstyle 0}|}\; \;\;\; 
< 1.5 \times 10^{-7} \;.
\end{equation}

As to be explicitly illustrated below [and discussed with expression 
(\ref{1st+})], this kind of contribution is insensitive to the  $\tan\!\zb$. 
Recall that imposing a sub-eV bound for all neutrino mass terms obtainable from
any RPV parameters (see, for example, Ref.\cite{ok}) gives 
$\frac{|\mu_i|}{|\mu_{\scriptscriptstyle 0}|} \, \cos\!\zb \lsim 10^{-6}$. A
simple estimate of a 1-loop neutrino mass diagram from two $\lambda$-type 
couplings give the corresponding bounds
$
|\lambda_{\scriptscriptstyle 121} \, \lambda_{\scriptscriptstyle 212}| \;,
|\lambda_{\scriptscriptstyle 321} \, \lambda_{\scriptscriptstyle 312}| \; 
\lsim 0.015
$,
and $|\lambda_{\scriptscriptstyle 212}| \lsim 0.008$. These bounds are 
not very strong at all as factors of $m_\mu$ and $m_e$ are involved in the
neutrino mass diagrams. Moreover, $\lambda_{\scriptscriptstyle 212}$ is the only 
parameter that is capable of giving rise to a 1-loop diagram just on its own,
hence with a bound on itself alone. The other four $\lambda$ couplings are otherwise 
bounded by $0.04$\cite{lambda}, from charged current processes and $\tau$ decays.
The $\mu_i$ parameters typically have no strong bound apart from the one due to
neutrino masses (see Ref.\cite{k} for details), which is suggested but 
not mandated from the result of the super-Kamiokande experiment\cite{sK}. 
Hence, we can see that the bound we obtained here from $\mu\to e\, \gamma$ 
is very important, especially in the large $\tan\!\zb$ region where the
neutrino mass bound is weakened. Further improvement in the 
$\mu\to e \gamma$ experiment is capable of giving the best bound on the 
RPV parameters, or discovering the signal of th R-parity violation.

\subsection{\boldmath\protect 
The  $\mu_{\scriptscriptstyle 1}^*\, \mu_{\scriptscriptstyle 2}$ contribution.}
The contribution is dominated by the 
${\cal C}^{\scriptscriptstyle R}_{1nm}\, {\cal C}^{\scriptscriptstyle L^*}_{2nm}$ 
term of $A^{\!\scriptscriptstyle R}_2$ as depicted explicitly in Fig.~2
[{\it cf.}  expression (\ref{2murl1})].  The result obviously contains a 
``muon Yukawa" suppression. The same story goes for the contribution from the 
${\cal C}_{\!\scriptscriptstyle 1nm}^{\scriptscriptstyle L} \,
{\cal C}_{\!\scriptscriptstyle 2nm}^{\scriptscriptstyle L^*}$
part of $A_2^{\!\scriptscriptstyle L}$, depicted in Fig.4a 
[{\it cf.} expression (\ref{ext12})]. The latter is typically 
numerically smaller. We show contours of $B(\mu\to e \,\gamma)$ in the real
$(\mu_{\scriptscriptstyle 1}, \mu_{\scriptscriptstyle 2})$
plane in Fig.~\ref{fig6}.  The 90\% C.L. upper bound on 
$\mu_{\scriptscriptstyle 1}^*\, \mu_{\scriptscriptstyle 2}$, normalized by 
$|\mu_{\scriptscriptstyle 0}|^2$ ($|\mu_{\scriptscriptstyle 0}|=100\;{\rm GeV}$
here), is
\begin{equation}
\frac{|\mu_{\scriptscriptstyle 1}^*\, \mu_{\scriptscriptstyle 2}|}
{|\mu_{\scriptscriptstyle 0}|^2} < 0.53 \times 10 ^{-4} \;.
\end{equation}
Note that the bound weakens roughly by a factor of 
${m_{\!\mu} \over \mu_{\scriptscriptstyle 0}}$ compared with
$\frac{|{\mu_{\scriptscriptstyle 3}^*}\,{\lambda_{\scriptscriptstyle 321}}|}
{|\mu_{\scriptscriptstyle 0}|}$, 
as expected. Though the bound looks weak compared with the sub-eV neutrino
mass bound discussed above, it is still significant when compared with most
of the other bounds on the $\mu_{\scriptscriptstyle 1}$ and
$\mu_{\scriptscriptstyle 2}$ parameters, especially in the region of
large $\tan\!\zb$\cite{k}. 

\subsection
{\boldmath\protect The contributions from two $\lambda$-type couplings.}
Here we have naively two classes of contributions, a class of
$\lambda^*\lambda$ diagrams and a class of $\lambda\lambda$
or $\lambda^*\lambda^*$ diagrams. The first class needs a chirality flip on the
external muon line. We have  a
$\lambda_{{\scriptscriptstyle 13}k}^*\, \lambda_{{\scriptscriptstyle 32}k}$ 
combination (for $k=1,2,3$) from the 
${\cal C}_{\!\scriptscriptstyle 1nm}^{\scriptscriptstyle R} \,
{\cal C}_{\!\scriptscriptstyle 2nm}^{\scriptscriptstyle R^*}$ part
of $A_2^{\!\scriptscriptstyle R}$ [{\it cf.} expression (\ref{2lam})], and a
$\lambda_{ij{\scriptscriptstyle 2}}^*\, \lambda_{{\scriptscriptstyle ji}1}$
combination (for $ij=12,13,23$) from the 
${\cal C}_{\!\scriptscriptstyle 1nm}^{\scriptscriptstyle L} \,
{\cal C}_{\!\scriptscriptstyle 2nm}^{\scriptscriptstyle L^*}$ part
of $A_2^{\!\scriptscriptstyle L}$. The first group must have a
$l_k^{\scriptscriptstyle +}$ running in the loop, while the second admits
both a $l_i^{\!\!\mbox{ -}}$ and a $l_j^{\!\!\mbox{ -}}$ and hence gives a twice
stronger result (with roughly degenerate $L$ and $R$ sleptons). The 
bounds obtained on appropriate combinations of $\lambda \lambda^*$ 
are given in Table~\ref{table2}, in which case we have deliberately 
used $\tan\!\beta=1$ (instead of $10$, in order keep the physical $L$- and $R$-handed 
slepton masses at about $100\,\mbox{GeV}$) 
to show the exact agreement with Ref.\cite{CH}. Note again the numerical bounds
are roughly a factor of $10^{-3}$ weaker than the $\mu^*\,\lambda$-type bound,
as a result of the $m_\mu$ factor.

On the other hand,  a  $\lambda\lambda$ or $\lambda^*\lambda^*$ diagram 
requires no chirality flip outside the loop. In the $\lambda\lambda$ case,
for example, we have to pick a $\lambda_{h'{\scriptscriptstyle 2}k'}$ 
to get the required reduction in $L_\mu$ and a $\lambda_{hk{\scriptscriptstyle 1}}$
to get the increase in $L_e$. It is easy to see then that it is impossible
to choose a combination of $\lambda\lambda$ such that  it does not 
cause further changes in any of the lepton-flavor numbers,
unless more RPV parameters are involved.  A term of the latter case has been 
discussed analytically [{\it cf.} expression (\ref{maj})]. The same 
situation holds for $\lambda^* \lambda^*$.  We will not further investigate 
such contributions here.

\subsection{\boldmath\protect The contributions involving the ${B_i}$ parameters}
We have discussed above the implication of the tadpole equations 
relating the $B_i$'s to the $\mu_i$'s.  Our numerical strategy
isolates terms explicitly proportional to a $B_i$ or a $\mu_i$. Here we 
discuss the contributions involving the $B_i$'s.  First, there are combinations
${B_{k}^*}\,{\lambda_{k\scriptscriptstyle 21}}$ and
${B_{k}}\,{\lambda_{k\scriptscriptstyle 12}^*}$. For 
${B_{k}^*}\,{\lambda_{k\scriptscriptstyle 21}}$ an illustrative contribution 
has been discussed analytically in the previous section
[{\it cf.} expressions (\ref{Blam}) and (\ref{Blam2})]. 
The bounds obtained on these combinations are shown in Table~\ref{table2}. 
Contours of $B(\mu\to e\, \gamma)$ in the real
$(B_{3},\lambda_{\scriptscriptstyle 321} )$ plane are shown in  Fig.~\ref{fig7}. 
Understanding the result analytically is more complicated here. Our analysis 
does suggest more than one factor of  Yukawa suppression, hence the weakness of 
the bound obtained. Numerically, the part with external chirality flip 
from a ${\cal C}_{\!\scriptscriptstyle 1nm}^{\scriptscriptstyle L} \,
{\cal C}_{\!\scriptscriptstyle 2nm}^{\scriptscriptstyle L^*}$ contribution
[{\it cf.} expression (\ref{Blam2})] can dominate over the part with the
chirality flip inside the loop [{\it cf.} expression (\ref{Blam})] depending on 
$\tan\beta$. The overall bound looks stronger than the analytical estimate. 
This is to be explained by the larger loop function $F_5$ with the much lighter
muon propagator compared with that of a chargino, and the $\tan\!\zb$ dependence.
However, the relation between a $B_k$ and a $\mu_k$ and
the weakness of the present result compared to the 
${\mu_{k}^*}\,{\lambda_{k\scriptscriptstyle 21}}$ or
${\mu_{k}}\,{\lambda_{k\scriptscriptstyle 12}^*}$ result suggests that the
$B^*\lambda$- or $B\lambda^*$- type contribution is really of less 
significance.

Next, we have the $B_1^*\, \mu_{\scriptscriptstyle 2}$ combination. Recall that
according to our strategy, this is probed with $\mu_{\scriptscriptstyle 1}$
and $B_2$ set to zero. It clearly has a ``muon Yukawa" suppression 
[{\it cf.} expression (\ref{2murl2})]. Numerically, the bound is similar to 
that on $\mu_{\scriptscriptstyle 1}^*\, \mu_{\scriptscriptstyle 2}$. The number
shown in Table~\ref{table2}  actually looks better than the corresponding
$\mu_{\scriptscriptstyle 1}^*\, \mu_{\scriptscriptstyle 2}$ number. However,
$B_1$ comes in with a $\tan\!\zb$ dependence. 
Therefore,  $\frac{|\mu_{\scriptscriptstyle 1}|}{|\mu_{\scriptscriptstyle 0}|}$
should be compared with 
$\frac{|B_1 \, \tan\!\zb|}{|\mu_{\scriptscriptstyle 0}|^2}$,
hence explaining the difference. Again, we give contours of $B(\mu\to e \,\gamma)$ 
in the real $(B_{1},\mu_{\scriptscriptstyle 2})$ plane, in  Fig.~\ref{fig8}. 

The analogous contribution of the type 
coming from $\mu_{\scriptscriptstyle 1}^*\, B_2$ has a ``electron Yukawa"
suppression. Numerically, it is confirmed to be much smaller. Likewise, we find
no important contribution from an explicit $B_1^*\, B_2$ combination. 
In fact, one obviously cannot make a simple $\mu \to e\,\gamma$
diagram with a $B_1^*$ and a $B_2$ RPV insertions.

\subsection{Parameter Variations}
In this section, we illustrate the effects of varying the 
input parameters on the bounds, using  
$|\mu_{\scriptscriptstyle 3}^*\,\lambda_{k\scriptscriptstyle 21}|$
and  $|\mu_{\scriptscriptstyle 1}^*\,\mu_{\scriptscriptstyle 2}|$ as examples.
The results are summarized in Table~\ref{table3}.
Basically, the effects of varying the mass parameters 
($\mu_{\scriptscriptstyle 0}, M_{\scriptscriptstyle 1}, 
\tilde{m}^2_{\!\scriptscriptstyle L}, \tilde{m}^2_{\!\scriptscriptstyle E}$) 
reflect on what particles are inside the loop of the dominant diagrams.
In the case of 
$|\mu_{\scriptscriptstyle 3}^*\,\lambda_{\scriptscriptstyle 321}|$, 
the dominant diagram involves mainly the 
$\tilde{l}_{\scriptscriptstyle 2}^{\scriptscriptstyle 0}$,
while the $|\mu_{\scriptscriptstyle 1}^*\,\mu_{\scriptscriptstyle 2}|$ case
involves the mixing between 
$\tilde{l}_{\scriptscriptstyle 2}^{\scriptscriptstyle 0}$
and $\tilde{l}_{\scriptscriptstyle 1}^{\scriptscriptstyle 0}$.
Therefore, varying $\tilde{m}^2_{\!\scriptscriptstyle E}$ does not have much 
effect on the bounds while varying the corresponding entries in 
$\tilde{m}^2_{\!\scriptscriptstyle L}$ changes the bounds significantly.  
Obviously, a large (relevant) scalar mass suppresses 
the $\mu \to e\,\gamma$ amplitude and thus weakens the bounds [see parts
(iii) and (iv)].

Increasing $\mu_{\scriptscriptstyle 0}$ and 
$M_{\scriptscriptstyle 1}={1\over 2}\,M_{\scriptscriptstyle 2}$ essentially 
increases the chargino and neutralino masses [see parts (i) and (ii)]. 
Here, the variation of the bound 
is more complicated.  For illustration, we show more details of the 
$\mu_{\scriptscriptstyle 0}$ variations in part (i). Taking
$|\mu_{\scriptscriptstyle 3}^*\,\lambda_{\scriptscriptstyle 321}|$
as an example, the bound is most stringent for small $|\mu_{\scriptscriptstyle 0}|$.
However, for negative $\mu_{\scriptscriptstyle 0}$, apart from the general weakening 
trend as  $|\mu_{\scriptscriptstyle 0}|$ increases, there is an extra structure in
the dominating chargino contribution, namely, there is a dip at 
$\mu_{\scriptscriptstyle 0} = - M_{\!\scriptscriptstyle 2}\, \tan\!\beta$,
where it essentially vanishes. This is a special feature of the type of RPV 
contribution also observed in the similar contribution to neutron electric dipole 
moment\cite{as6}. For the relatively large value of $\tan\!\beta$, used here, 
however, this is already 
well inside the large $|\mu_{\scriptscriptstyle 0}|$ region where 
the pure gauge loop contribution becomes dominant. The latter case generally
happens when the bino mass $M_{\scriptscriptstyle 1}$ is small relative to
$|\mu_{\scriptscriptstyle 0}|$. Note that the pure gauge loop contribution
is independent of $\mu_{\scriptscriptstyle 0}$. Apart from the dip mentioned, 
the dominant chargino contribution does decrease with increasing  
$|\mu_{\scriptscriptstyle 0}|$, as shown in column~1 of Table~\ref{table3}.

As for the  $|\mu_{\scriptscriptstyle 1}^*\,\mu_{\scriptscriptstyle 2}|$ case,
the variation of the bound with $|\mu_{\scriptscriptstyle 0}|$ is more complicated.
Unlike the previous case, the contribution has, analytically, an explicit chargino 
mass dependence [as shown by comparing expression (\ref{2murl1})  with expression 
(\ref{1st+})].  Hence, we do not expect a simple weakening of the result as  
$|\mu_{\scriptscriptstyle 0}|$ increases. Another important issue here is that
there is more than one important piece of contribution of this type in most regions 
of the parameter space. For example, if we stick to setting $B_1=B_2=0$ as we do to
obtain the numbers given in column~2 of Table~\ref{table3}, we still have
the pieces corresponding to Fig.~2 and Fig.~4a [{\it cf.} expressions 
(\ref{2murl1}) and (\ref{ext12})] in interference with one another. 
Besides, at very large $|\mu_{\scriptscriptstyle 0}|$ (and relatively small 
$M_{\!\scriptscriptstyle 1}$)  the pure gauge-gauge term of
the neutralino-like contribution is increasing and dominates over the 
chargino-like contribution.

In addition, we have shown in column~3 of the Table~\ref{table3}, an interesting
case with $\tilde{m}^2_{\!{\scriptscriptstyle L}_{\!{\scriptscriptstyle 0}i}}=0$
instead. Then the $\mu_1^*\,\mu_{\scriptscriptstyle 2}$- and 
$B_1^*\,\mu_{\scriptscriptstyle 2}$-type contributions
exist simultaneously.  As given in Eq.(\ref{tp3}), we then have 
$B_1\,\tan\!\zb = \mu_{\scriptscriptstyle 0}^*\,\mu_{\scriptscriptstyle 1}$,
and the contribution of $B_1^*\,\mu_{\scriptscriptstyle 2}$ type such as that
given by expression (\ref{2murl2}) comes along and interferes with that of
the  $\mu_{\scriptscriptstyle 1}^*\,\mu_{\scriptscriptstyle 2}$. 
Our result in Table~\ref{table2} suggests that the 
$B_1^*\,\mu_{\scriptscriptstyle 2}$ 
type contribution is smaller than the 
$\mu_{\scriptscriptstyle 1}^*\,\mu_{\scriptscriptstyle 2}$ type
but are of the same order; a destructive interference of the two parts then weakens 
the overall result. Note that our analytical discussions in the previous section
[{\it cf.} expressions (\ref{2murl1}) and (\ref{2murl2})] indicate that the terms
come in with a different sign. With larger $|\mu_{\scriptscriptstyle 0}|$,
the bounds shown in column~3 of the table are much weaker, because of a 
stronger cancellation as the $B_1^*\,\mu_{\scriptscriptstyle 2}$ term increases.
This is mainly a result of the increase in the $B_1$ value (for fixed 
$\mu_{\scriptscriptstyle 1}$), however, 
the contribution from the term has more complicated dependence on the 
$|\mu_{\scriptscriptstyle 0}|$. Similarly, the same increase
in cancellation with larger $\tilde{m}^2_{\!{\scriptscriptstyle L}}$, as shown
in the table can be understood from comparing the two terms.
The results shown in column~3, hence, further illustrate the importance of the
tadpole condition given by Eq.(\ref{tp3}) emphasized throughout the paper.

Finally, we comment briefly on the $\tan\!\beta$ dependence of the results,
also illustrated in part (v) of Table~\ref{table3}. From the table  
we can see that varying $\tan\!\beta$ has only a little effect on 
$|\mu_{\scriptscriptstyle 3}^*\,\lambda_{\scriptscriptstyle 321}|$ but 
a rather significant effect on 
$|\mu_{\scriptscriptstyle 1}^*\,\mu_{\scriptscriptstyle 2}|$. The lack of
sensitivity to $\tan\!\beta$ in the former case has been suggested in our
analytical discussion [{\it cf.} expression (\ref{1st+})] and further 
confirmed over the range of the $\tan\!\beta$ value. In the latter case, the 
numerical result shows that the bound is strengthened by a factor of
$\cos\!\beta$. This simply illustrates the ${1\over \cos\!\beta}$ dependence
of the Yukawa coupling ${y_{\!\scriptscriptstyle e_2}}$.

\section{Conclusion}
In this paper, we have performed a comprehensive study on the radiative decay 
of muon ($\mu\to e \,\gamma$) in the framework of the generic supersymmetric 
standard model (without R parity).  We have identified a few combinations of a 
minimal number of RPV couplings  contributing to the decay.
Among them the most interesting are
$\mu_k^*\,\lambda_{k\scriptscriptstyle 21}$ and
$\mu_k\,\lambda_{k\scriptscriptstyle 12}^*$.  The upper bound on the 
combinations obtained from the experimental limit on $\mu\to e\,\gamma$ is
\[
\frac{|{\mu_{k}^*}\,{\lambda_{k\scriptscriptstyle 21}}|}
{|\mu_{\scriptscriptstyle 0}|}\;,
\frac{|{\mu_{k}}\,{\lambda_{k\scriptscriptstyle 12}^*}|}
{|\mu_{\scriptscriptstyle 0}|}\;
<
1.5 \times 10^{-7} \;,
\]
which is as stringent as the ones that can be obtained from the constraint of
sub-eV neutrino masses. Note that different combinations of RPV parameters
are involved in the generation of neutrino masses though. 
Furthermore, our result, in contrast to a similar result given 
in Ref.\cite{cch2}, has little sensitivity to $\tan\!\beta$. 

Another combination, $\mu_{\scriptscriptstyle 1}^* \,\mu_{\scriptscriptstyle 2}$,
contributing to $\mu \to e\,\gamma$ involves the $LL$-slepton mixing.  This 
contribution is identified for the first time. The upper bound on 
the combination obtained from the experimental limit on $\mu\to e\,\gamma$
could be important, especially in the region of large $\tan\!\zb$.
We have also discussed the related role of the soft SUSY breaking 
$B_i$ parameters in the process. 

Before closing we summarize the following important points :

(i) The combinations $\mu_{k}^* \,\lambda_{k\scriptscriptstyle 21}$
and $\mu_{\scriptscriptstyle k} \,\lambda_{k\scriptscriptstyle 12}^*$
participate directly in the $LR$-slepton mixings, while 
$\mu_{i}^* \,\mu_{j}$ participate in the $LL$-slepton mixings. The combinations also 
highlight the major $\mu\to e\, \gamma$ contributions. Under our formulation 
(SVP), the only RPV soft SUSY breaking parameters that contribute to 
(tree-level) slepton masses are the $B_i$'s and the
$\tilde{m}^2_{\!{\scriptscriptstyle L}_{\!{\scriptscriptstyle 0}i}}$'s.

(ii) In relation to the RPV parameters (under SVP), the tadpole equations 
say that $\{ \mu_{i}, B_{i}, 
\tilde{m}^2_{\!{\scriptscriptstyle L}_{\!{\scriptscriptstyle 0}i}} \}$
for each $i$ are not independent.  This fact is often overlooked in the literature.  
In our study, we single out the contribution from $\mu_i$ or $B_{i}$ with a matching
nonzero $\tilde{m}^2_{\!{\scriptscriptstyle L}_{\!{\scriptscriptstyle 0}i}}$. 
We have also illustrated a case with
$\tilde{m}^2_{\!{\scriptscriptstyle L}_{\!{\scriptscriptstyle 0}i}}=0$,
and thus combined the effects of both $\mu_{i}$ and $B_{i}$. This case 
shows that there could be strong cancellations between them.  Hence,
the overall contribution from the bilinear RPV parameters only is tied up with 
how the tadpole equations among $\{ \mu_{i}, B_{i}, 
\tilde{m}^2_{\!{\scriptscriptstyle L}_{\!{\scriptscriptstyle 0}i}} \}$'s 
are chosen to be satisfied. 

\acknowledgements
This work was supported in part by the National Center for Theoretical Science 
under a grant from the National Science Council of Taiwan R.O.C., and in part
by Academia Sinica.

\newpage
\noindent
{\bf Table captions :}\\[.1in]
Table I --- 
Basic input SUSY parameters for the numerical results presented. These 
values are adopted unless otherwise specified. Note that 
$\tilde{m}^2_{\!{\scriptscriptstyle L}_{\!{\scriptscriptstyle 00}}}$ corresponds 
to the soft mass square for $H_d$, in the MSSM language. Moreover, soft masses 
for $H_u$ and  $B_0$ ($\sim$ MSSM soft SUSY breaking $B$ term) are not used as 
inputs, but are determined from 
$\tilde{m}^2_{\!{\scriptscriptstyle L}_{\!{\scriptscriptstyle 00}}}$
and $\mu_{\scriptscriptstyle 0}$ ($\sim$ MSSM $\mu$ term) through the tadpole
equations for correct electroweak symmetry breaking.\\[.1in]
Table II ---
Summary of bounds on various 
combinations of two R-parity violating parameters, 
normalized by $|\mu_{\scriptscriptstyle 0}|=100$ GeV 
where appropriate, due to the 
experimental limit $B(\mu\to e\,\gamma) < 1.2\times 10^{-11}$ at 90\% C.L.
The input parameters are as in Table~\ref{table1}, except for the limits on 
$|\lambda\lambda^*|$ combinations in which case we have used $\tan\!\beta=1$ 
(as explained in the text).\\[.1in]
Table III --- 
Effects of parameter variations of interest, on the bounds of
${|{\mu_{\scriptscriptstyle 3}^*}\,{\lambda_{\scriptscriptstyle 321}}|}
\cdot {(100\,\mbox{GeV})^{-1}}$	and
${|\mu_{\scriptscriptstyle 1}^*\, \mu_{\scriptscriptstyle 2}|}
\cdot {(100\,\mbox{GeV})^{-2}}$. Note that the fixed mass scale of $100\,\mbox{GeV}$ is used for normalization to extract numerical bounds.
The tadpole condition is chosen with $B_i=0$ in the second column of the bounds
while $\tilde{m}^2_{L_{0i}}=0$ is chosen in the last column.
\\

\newpage

\noindent
{\bf Figure captions :}\\[.1in]
Fig. 1 --- The R-parity violating chargino-like loop diagram. \\
Fig. 2 --- Diagram with charged gaugino/higgsino mixing.\\
Fig. 3 --- A diagram involving Majorana-like scalar mass insertion. 3a/ The diagram.
3b/ The seesaw origin of Majorana-like scalar masses for the ``sneutrinos" explicitly
illustrated. \\
Fig. 4 --- Illustrative diagrams with chirality flip on the external muon line.
4a/. A charged gaugino loop. 4b/ A $g-\lambda$ neutralino-like loop.
\\
Fig. 5 --- Contours of $B(\mu \to e \,\gamma)$ in the (real) plane of  
(${\mu_{\scriptscriptstyle 3}}, \lambda_{\scriptscriptstyle 321}$).
The 90\% C.L. allowed region is shaded. \\
Fig. 6 --- Contours of $B(\mu \to e \,\gamma)$ in the (real) plane of  
(${\mu_{\scriptscriptstyle 1}},{\mu_{\scriptscriptstyle 2}} $).
The 90\% C.L. allowed region is shaded.
Note that the approximation we used for the external lepton
lines is less applicable at the right and top ends of the plot where
the result should be read with caution.
Fig. 7 --- Contours of $B(\mu \to e \,\gamma)$ in the (real) plane of  
(${B_{3}}, \lambda_{\scriptscriptstyle 321}$).
The 90\% C.L. allowed region is shaded. \\
Fig. 8 --- Contours of $B(\mu \to e \,\gamma)$ in the (real) plane of  
($B_1,{\mu_{\scriptscriptstyle 2}} $).
The 90\% C.L. allowed region is shaded. 

\newpage

\begin{table}[bh]
\vspace*{.5in}
\caption{\small \label{table1}
Basic input SUSY parameters for the numerical results presented. These 
values are adopted unless otherwise specified. Note that 
$\tilde{m}^2_{\!{\scriptscriptstyle L}_{\!{\scriptscriptstyle 00}}}$ corresponds 
to the soft mass square for $H_d$, in the MSSM language. Moreover, soft masses 
for $H_u$ and  $B_0$ ($\sim$ MSSM soft SUSY breaking $B$ term) are not used as 
inputs, but are determined from 
$\tilde{m}^2_{\!{\scriptscriptstyle L}_{\!{\scriptscriptstyle 00}}}$
and $\mu_{\scriptscriptstyle 0}$ ($\sim$ MSSM $\mu$ term) through the tadpole
equations for correct electroweak symmetry breaking.}
\bigskip
\begin{tabular}{cccc}
$M_{\scriptscriptstyle 1}$ (GeV) & $M_{\scriptscriptstyle 2}$ (GeV)  & $\mu_{\scriptscriptstyle 0}$ (GeV)  & $\tan\!\beta$ \\
\hline 
100  & 200 & 100 & 10 \\
\hline
\hline
$\tilde{m}^2_{\!{\scriptscriptstyle L}}$ ($10^4$ GeV$^{2}$) &
$\tilde{m}^2_{\!{\scriptscriptstyle E}}$ ($10^4$ GeV$^{2}$) & 
$A_e$ (GeV) & \\
\hline
diag$\{2,1,1,1\}$ &
diag$\{1,1,1\}$ & 100 & \\
\end{tabular}
\end{table}

\vspace*{1in}

\begin{table}[bh]
\caption{\small \label{table2}
Summary of bounds on various 
combinations of two R-parity violating parameters, 
normalized by $|\mu_{\scriptscriptstyle 0}|=100$ GeV 
where appropriate, due to the 
experimental limit $B(\mu\to e\,\gamma) < 1.2\times 10^{-11}$ at 90\% C.L.
The input parameters are as in Table~\ref{table1}, except for the limits on 
$|\lambda\lambda^*|$ combinations in which case we have used $\tan\!\beta=1$ 
(as explained in the text).
}
\bigskip
\begin{tabular}{ll}
\ \ 
$\frac{|{\mu_{\scriptscriptstyle 3}^*}\,{\lambda_{\scriptscriptstyle 321}}|}
{|\mu_{\scriptscriptstyle 0}|}\;, \;\;\;
\frac{|{\mu_{\scriptscriptstyle 1}^*}\,{\lambda_{\scriptscriptstyle 121}}|}
{|\mu_{\scriptscriptstyle 0}|}\;, \;\;\;
\frac{|{\mu_{\scriptscriptstyle 3}}\,{\lambda_{\scriptscriptstyle 312}^*}|}
{|\mu_{\scriptscriptstyle 0}|}\;, \;\;\; 
\mbox{or} \;\;\;
\frac{|{\mu_{\scriptscriptstyle 2}}\,{\lambda_{\scriptscriptstyle 212}^*}|}
{|\mu_{\scriptscriptstyle 0}|}\; \;\;\;$ & 
$< 1.5 \times 10^{-7}$ \ \ 
\\ \hline 
\ \
$\frac{|\mu_{\scriptscriptstyle 1}^*\, \mu_{\scriptscriptstyle 2}|}
{|\mu_{\scriptscriptstyle 0}|^2}$	&	$ < 0.53 \times 10 ^{-4}$ \ \
\\ \hline
\ \
$|\lambda_{\scriptscriptstyle 321} \lambda^*_{\scriptscriptstyle 131}|\;, \;\;\;
|\lambda_{\scriptscriptstyle 322} \lambda^*_{\scriptscriptstyle 132}|\;, \;\;\; 
\mbox{or} \;\;\;
|\lambda_{\scriptscriptstyle 323} \lambda^*_{\scriptscriptstyle 133}|$      
& $<2.2 \times 10^{-4}$ \ \ 
\\ \hline
\ \
$|\lambda^*_{\scriptscriptstyle 132} \lambda_{\scriptscriptstyle 131}|\;, \;\;\;
|\lambda^*_{\scriptscriptstyle 122} \lambda_{\scriptscriptstyle 121}|\;, \;\;\; 
\mbox{or} \;\;\;
|\lambda^*_{\scriptscriptstyle 232} \lambda_{\scriptscriptstyle 231}|$      
& $<1.1 \times 10^{-4}$ \ \ 
\\ \hline
\ \  
$\frac{|B_{3}^*\,\lambda_{\scriptscriptstyle 321}|}{|\mu_{\scriptscriptstyle 0}|^2}
\;, \;\;\;
\frac{|B_{1}^*\,\lambda_{\scriptscriptstyle 121}|}{|\mu_{\scriptscriptstyle 0}|^2}
\;, \;\;\;
\frac{|B_{3}\,\lambda_{\scriptscriptstyle 312}^*|}{|\mu_{\scriptscriptstyle 0}|^2}
\;, \;\;\; \mbox{or} \;\;\;
\frac{|B_{2}\,\lambda_{\scriptscriptstyle 211}^*|}{|\mu_{\scriptscriptstyle 0}|^2}$
&  $<2.0\times 10^{-3}$ \ \ 
\\ \hline
\ \ 
$\frac{|B_1^* \, \mu_{\scriptscriptstyle 2}|}{|\mu_{\scriptscriptstyle 0}|^3}$
 & $< 1.1\times 10^{-5}$ \ \
\end{tabular}
\end{table}

\newpage

\begin{table}[bh]
\caption{\small \label{table3}
Effects of parameter variations of interest, on the bounds of
${|{\mu_{\scriptscriptstyle 3}^*}\,{\lambda_{\scriptscriptstyle 321}}|}
\cdot {(100\,\mbox{GeV})^{-1}}$	and
${|\mu_{\scriptscriptstyle 1}^*\, \mu_{\scriptscriptstyle 2}|}
\cdot {(100\,\mbox{GeV})^{-2}}$. Note that the fixed mass scale of $100\,\mbox{GeV}$
is used for normalization to extract numerical bounds.
The tadpole condition is chosen with $B_i=0$ in the second column of the bounds
while $\tilde{m}^2_{L_{0i}}=0$ is chosen in the last column.
}
\bigskip
\begin{tabular}{llll}
{\large Parameter changes}  &  \multicolumn{3}{c}{\large Normalized numerical bounds} \\
 & \multicolumn{1}{c}{$\frac{|{\mu_{\scriptscriptstyle 3}^*}\,{\lambda_{\scriptscriptstyle 321}}|}{(100\; {\rm GeV})}$}
 & \multicolumn{1}{c}{$\frac{|\mu_{\scriptscriptstyle 1}^*\, \mu_{\scriptscriptstyle 2}|}{(100\; {\rm GeV})^2}$}
 & \multicolumn{1}{c}{$\frac{|\mu_{\scriptscriptstyle 1}^*\, \mu_{\scriptscriptstyle 2}|}{(100\; {\rm GeV})^2}$} \\
& & (with $B_{i} = 0$) &
(with $\tilde{m}^2_{\!{\scriptscriptstyle L}_{\!{\scriptscriptstyle 0}i}}=0$)
 \\
\hline
Original inputs of Table \ref{table1} &
  $<1.5 \times 10^{-7}$  & $<0.53 \times 10^{-4}$ & $< 2.3 \times 10^{-4}$  \\
\hline
(i) $\mu_{\scriptscriptstyle 0}=500$ GeV  
& $< 10\times 10^{-7}$ & $<0.80 \times 10^{-4}$ & $< 27\times 10^{-4}$ \\ 
{\ \ \ \ }$\mu_{\scriptscriptstyle 0}=250$ GeV  
& $< 5.8\times 10^{-7}$ & $<1.1 \times 10^{-4}$ & $< 14\times 10^{-4}$ \\
{\ \ \ \ }$\mu_{\scriptscriptstyle 0}=-100$ GeV  
& $< 2.0\times 10^{-7}$ & $<0.64 \times 10^{-4}$ & $<2.1\times 10^{-4}$ \\
{\ \ \ \ }$\mu_{\scriptscriptstyle 0}=-250$ GeV  
& $< 6.8\times 10^{-7}$ & $<1.1 \times 10^{-4}$ & $<8.8\times 10^{-4}$  \\
{\ \ \ \ }$\mu_{\scriptscriptstyle 0}=-500$ GeV  
& $< 11\times 10^{-7}$ & $<0.78 \times 10^{-4}$ & $<19\times 10^{-4}$ \\
\hline
(ii) $M_{\scriptscriptstyle 1}={1\over 2}{M_{\scriptscriptstyle 2}}=500$ GeV  
& $< 9.3\times 10^{-7}$ & $< 3.7\times 10^{-4}$ & $<16\times 10^{-4}$ \\
\hline
(iii) $\tilde{m}^2_{\!{\scriptscriptstyle L}}
=20000\times\; {\rm diag}\{1,1,1,1\}$ GeV$^{2}$  
   & $< 2.2\times 10^{-7}$ & $< 1.3\times 10^{-4}$ & $<18\times 10^{-4}$ \\
{\ \ \ \ }$\tilde{m}^2_{\!{\scriptscriptstyle L}}
={\rm diag}\{20000,1000^2,1000^2,1000^2\}$ GeV$^{2}$  
   & $< 2.4\times 10^{-6}$ & $< 840\times 10^{-4}$ & $<44\times 10^{-4}$ \\
\hline
(iv) $\tilde{m}^2_{\!{\scriptscriptstyle E}}
=20000\times \; {\rm diag}\{1,1,1\}$ GeV$^{2}$  
 & $< 1.5\times 10^{-7}$ & $< 0.55\times 10^{-4}$ & $<2.4\times 10^{-4}$  \\
{\ \ \ \ }$\tilde{m}^2_{\!{\scriptscriptstyle E}}
={\rm diag}\{1000^2,1000^2,1000^2\}$ GeV$^{2}$  
  & $< 1.5\times 10^{-7}$ & $< 0.59\times 10^{-4}$ & $<2.5\times 10^{-4}$  \\
\hline
(v) $\tilde{m}^2_{\!{\scriptscriptstyle L}}
={\rm diag}\{20000,500^2,500^2,500^2\}$ GeV$^{2}$  & & & \\
{\ \ \ \ }  $\tilde{m}^2_{\!{\scriptscriptstyle E}}
={\rm diag}\{500^2,500^2,500^2\}$ GeV$^{2}$  & & & \\
{\ \ \ \ \ \ }  $\mu_0=100$ GeV, $\tan\beta=2$ 
   & $< 0.87\times 10^{-6}$ & $< 500\times 10^{-4}$ & $<51\times 10^{-4}$ \\
{\ \ \ \ \ \ } \hspace{1in}                   $\tan\beta=10$ 
   & $< 1.1\times 10^{-6}$ & $< 67\times 10^{-4}$ & $<15\times 10^{-4}$ \\
{\ \ \ \ \ \ } \hspace{1in}                   $\tan\beta=50$ 
   & $< 1.2\times 10^{-6}$ & $< 13\times 10^{-4}$ & $<3.4\times 10^{-4}$ \\
{\ \ \ \ \ \ }  $\mu_0=250$ GeV, $\tan\beta=2$ 
   & $< 2.2\times 10^{-6}$ & $< 450\times 10^{-4}$ & $<150\times 10^{-4}$ \\
{\ \ \ \ \ \ } \hspace{1in}                    $\tan\beta=10$ 
   & $< 3.1\times 10^{-6}$ & $< 88\times 10^{-4}$ & $<50\times 10^{-4}$ \\
{\ \ \ \ \ \ } \hspace{1in}                    $\tan\beta=50$ 
   & $< 3.4\times 10^{-6}$ & $< 18\times 10^{-4}$ & $<12\times 10^{-4}$ \\
{\ \ \ \ \ \ }  $\mu_0=500$ GeV, $\tan\beta=2$ 
   & $< 5.1\times 10^{-6}$ & $< 660\times 10^{-4}$ & $< 450\times 10^{-4}$ \\
{\ \ \ \ \ \ } \hspace{1in}                     $\tan\beta=10$ 
   & $< 8.2\times 10^{-6}$ & $< 140\times 10^{-4}$ & $<177\times 10^{-4}$ \\
{\ \ \ \ \ \ } \hspace{1in}                    $\tan\beta=50$ 
   & $< 9.8\times 10^{-6}$ & $< 29\times 10^{-4}$ & $<46\times 10^{-4}$ 
\end{tabular}
\end{table}

\newpage

\begin{figure}
\vspace*{.5in}
\includegraphics{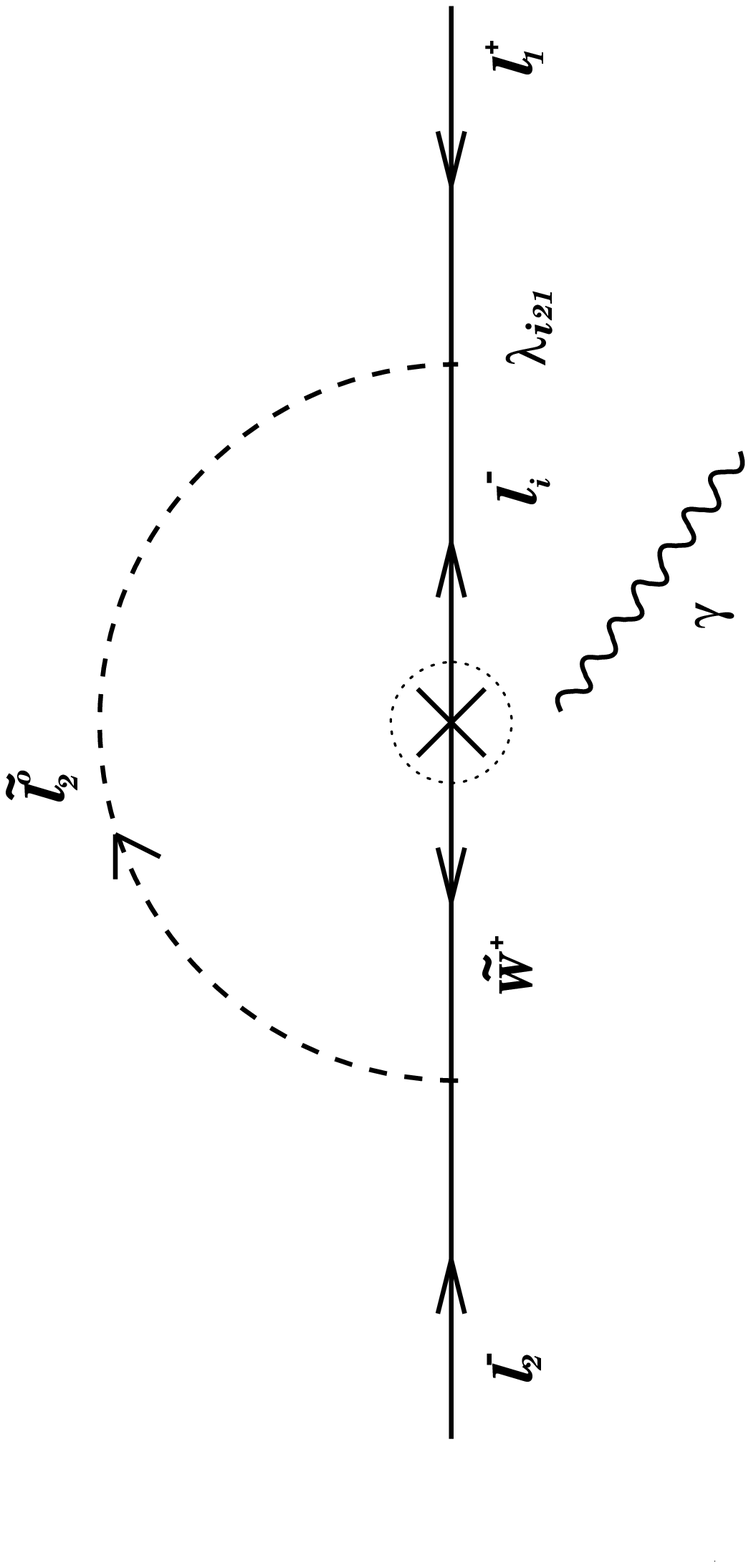}
\vspace*{3in}
\caption{The R-parity violating chargino-like loop diagram.}
\end{figure}

\vspace*{.8in}

\begin{figure}
\vspace*{.5in}
\includegraphics{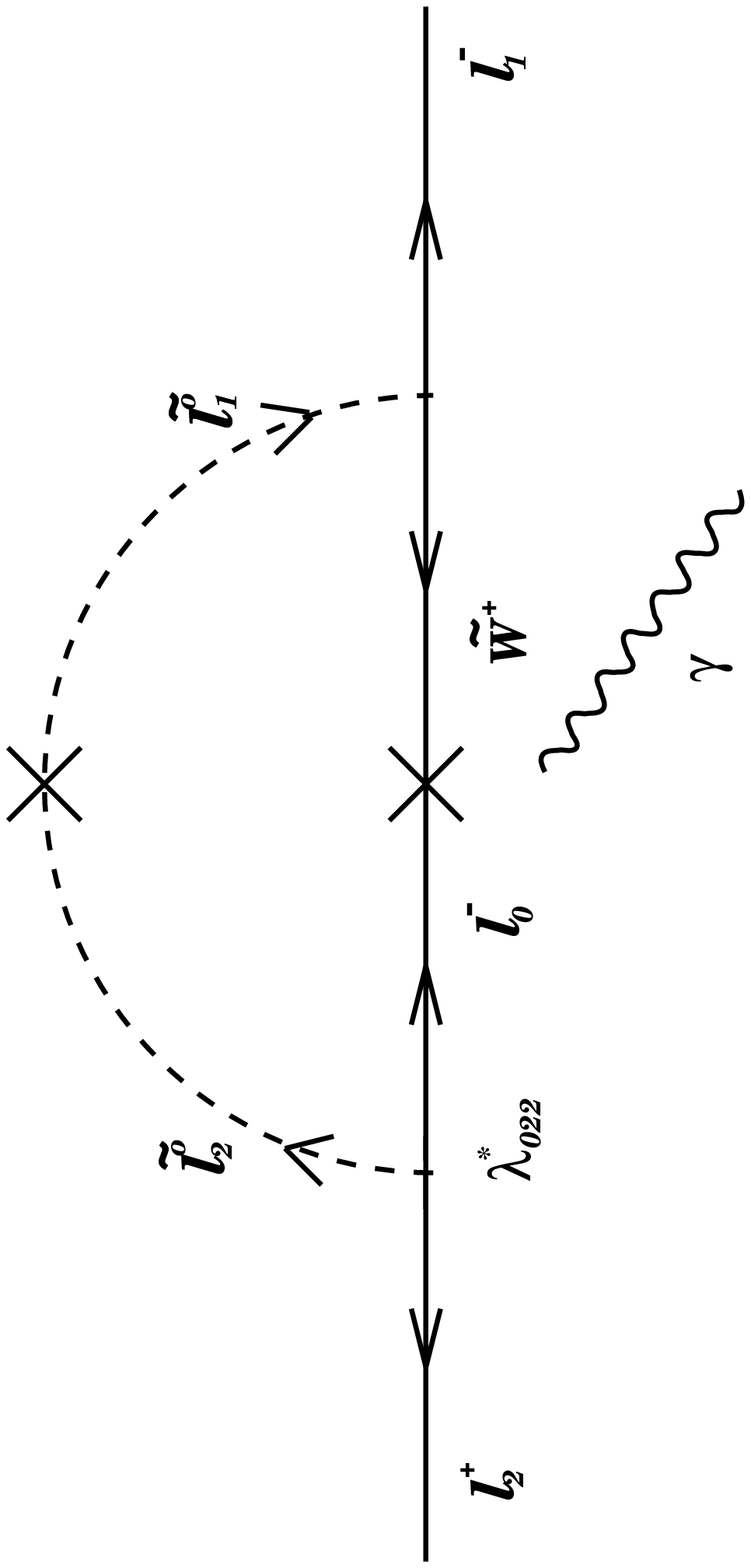}
\vspace*{3in}
\caption{Diagram with charged gaugino/higgsino mixing.}
\end{figure}

\eject

\begin{figure}
\vspace*{.5in}
\includegraphics{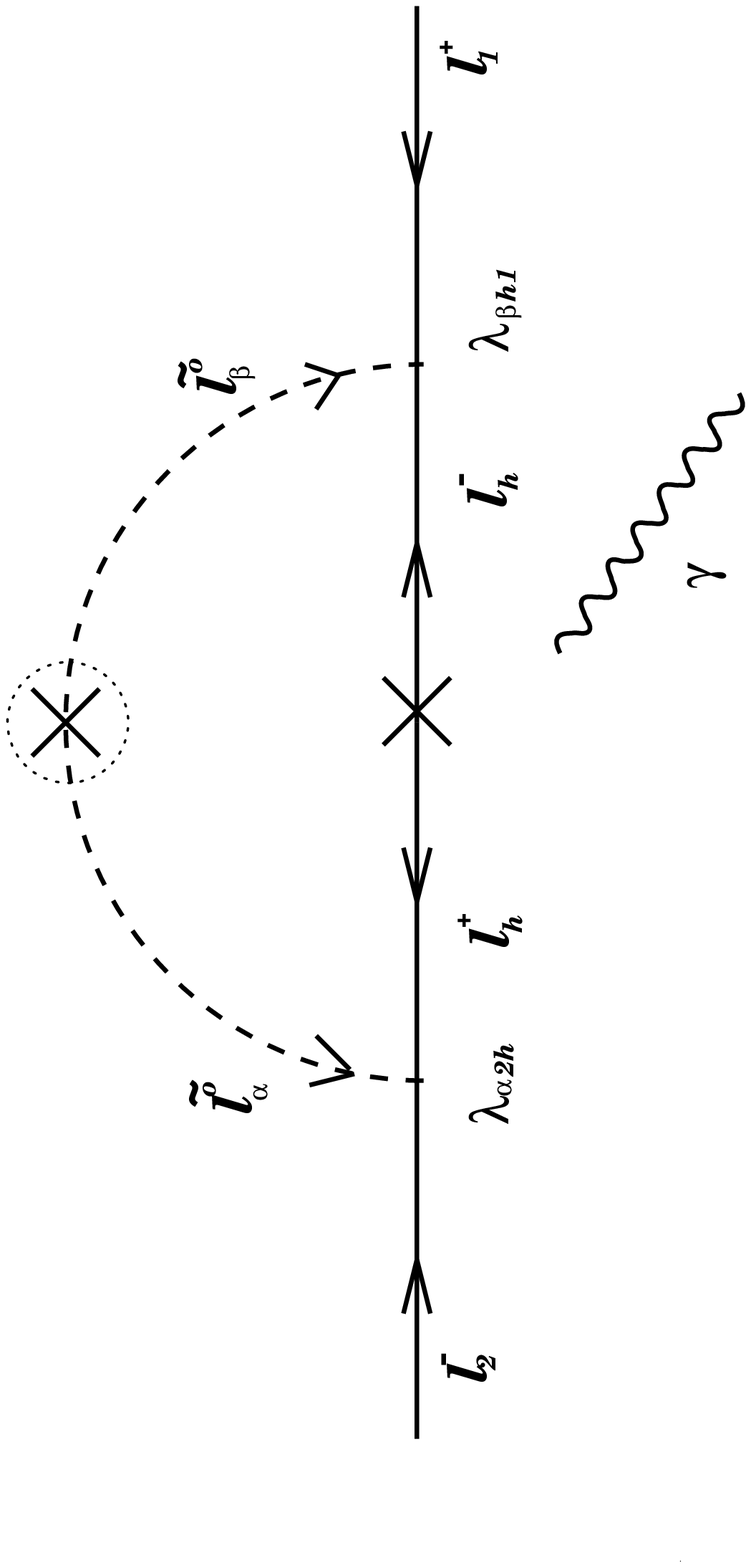}
\vspace*{3in}

\vspace*{.8in}

\vspace*{.5in}
\includegraphics{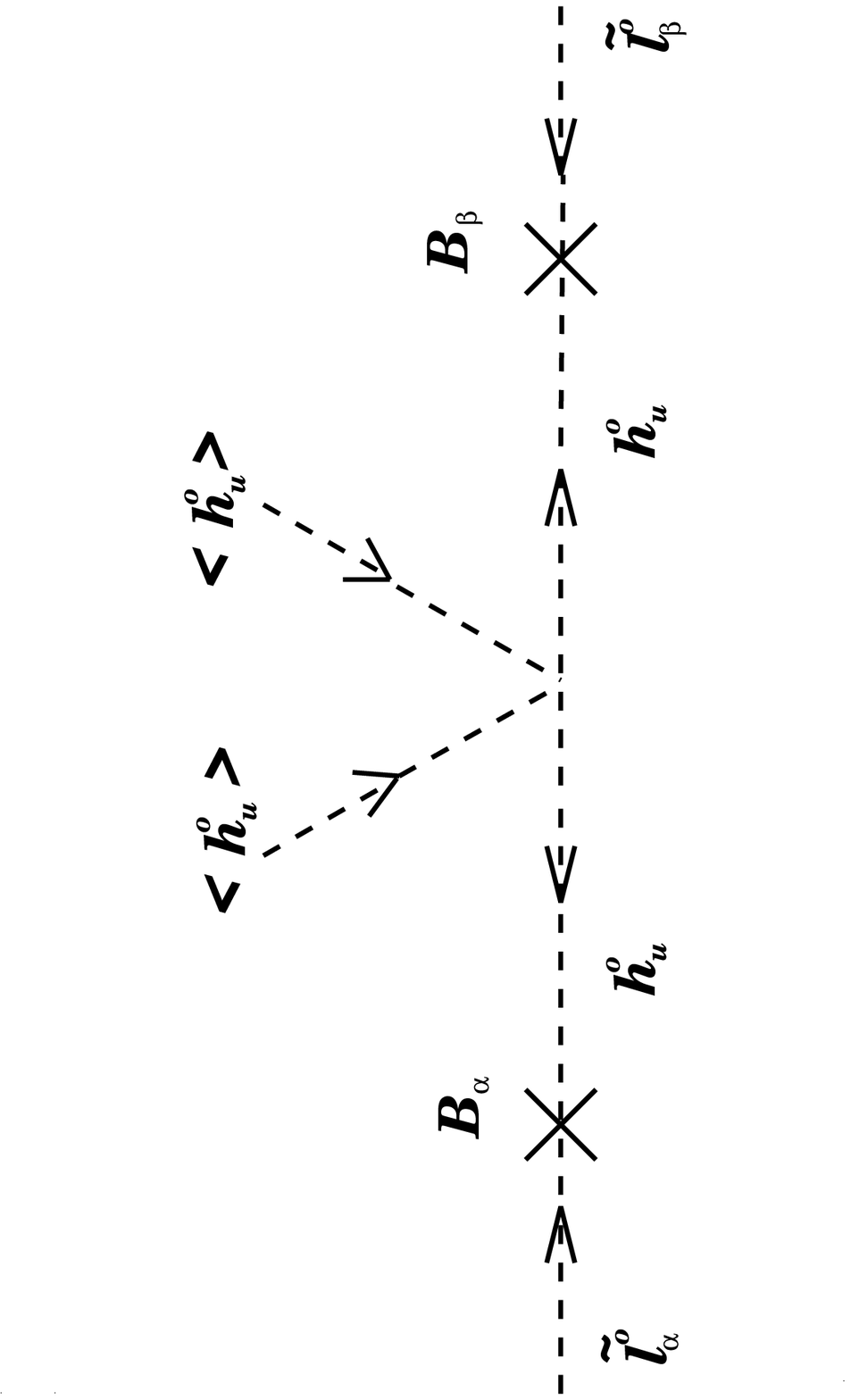}
\vspace*{3in}
\caption{A diagram involving Majorana-like scalar mass insertion. \ \
3a/ The diagram. \ \
3b/ The seesaw origin of Majorana-like scalar masses for the ``sneutrinos" explicitly
illustrated.}
\end{figure}

\eject
\begin{figure}
\vspace*{.5in}
\includegraphics{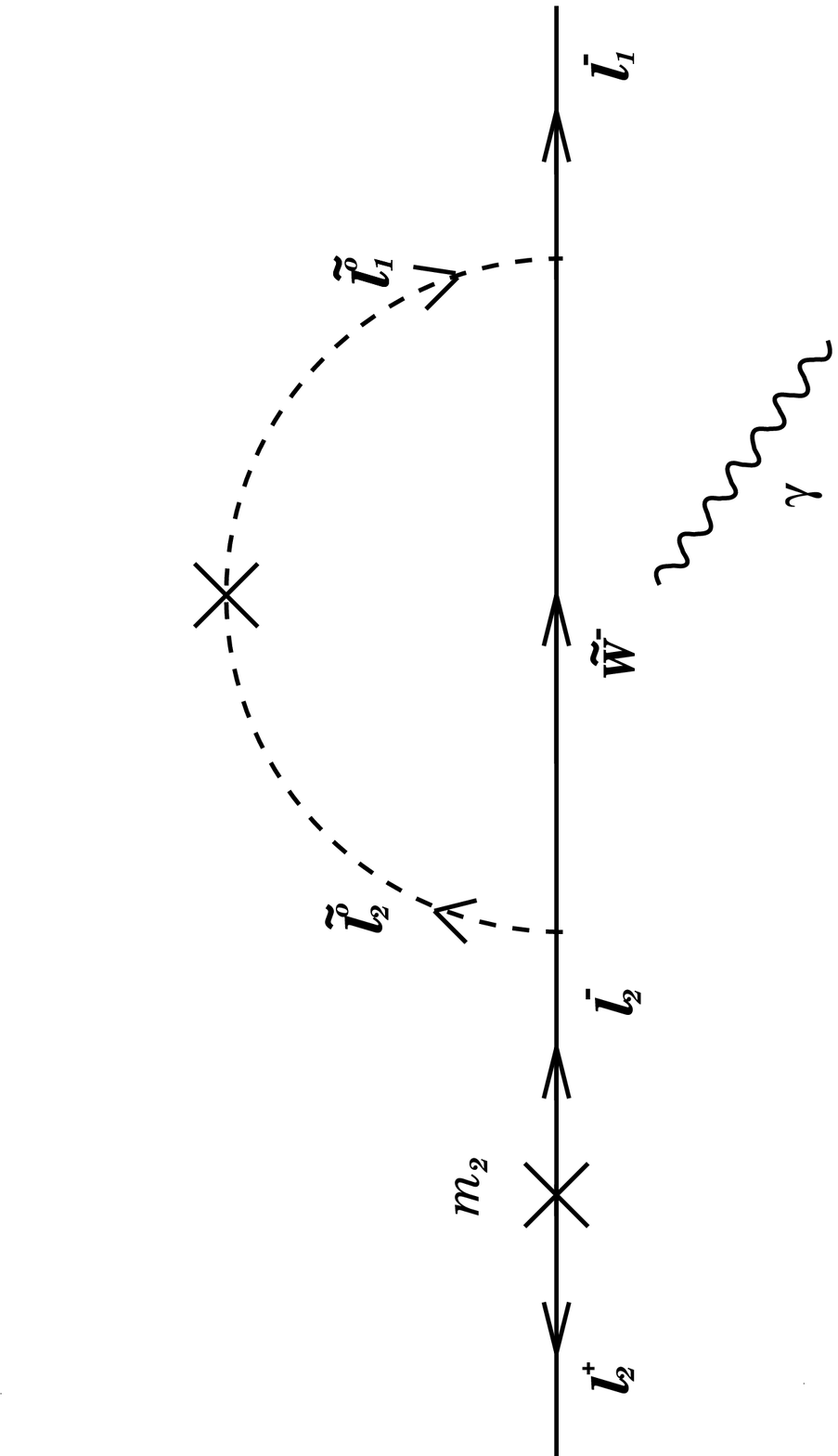}
\vspace*{3in}

\vspace*{.8in}

\vspace*{.5in}
\includegraphics{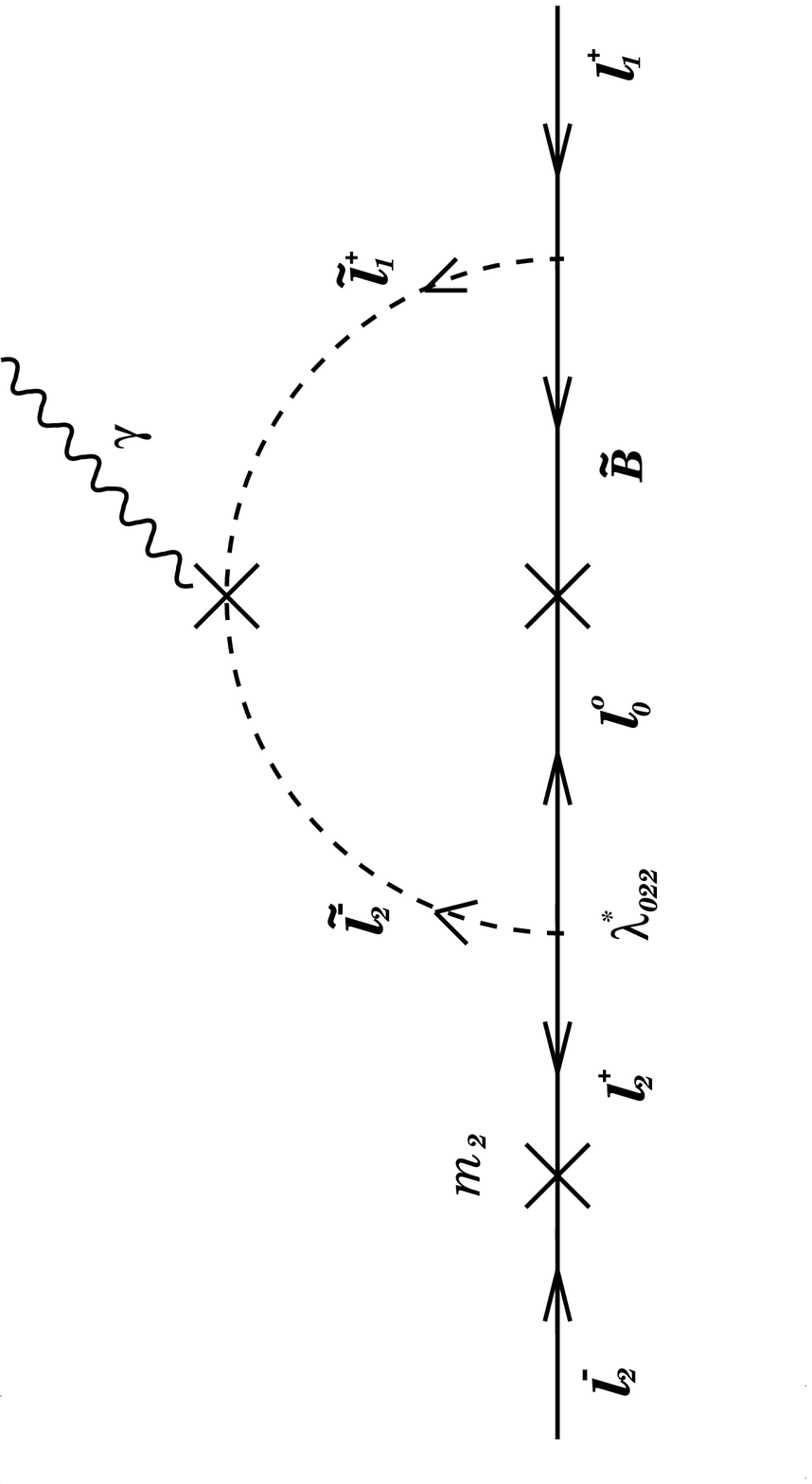}
\vspace*{3in}
\caption{\small Illustrative diagrams with chirality flip on the external 
muon line.\ \
4a/. A charged gaugino loop. \ \
4b/ A $g-\lambda$ neutralino-like loop.}
\end{figure}

\eject

\begin{figure}
\vspace*{3in}
\includegraphics{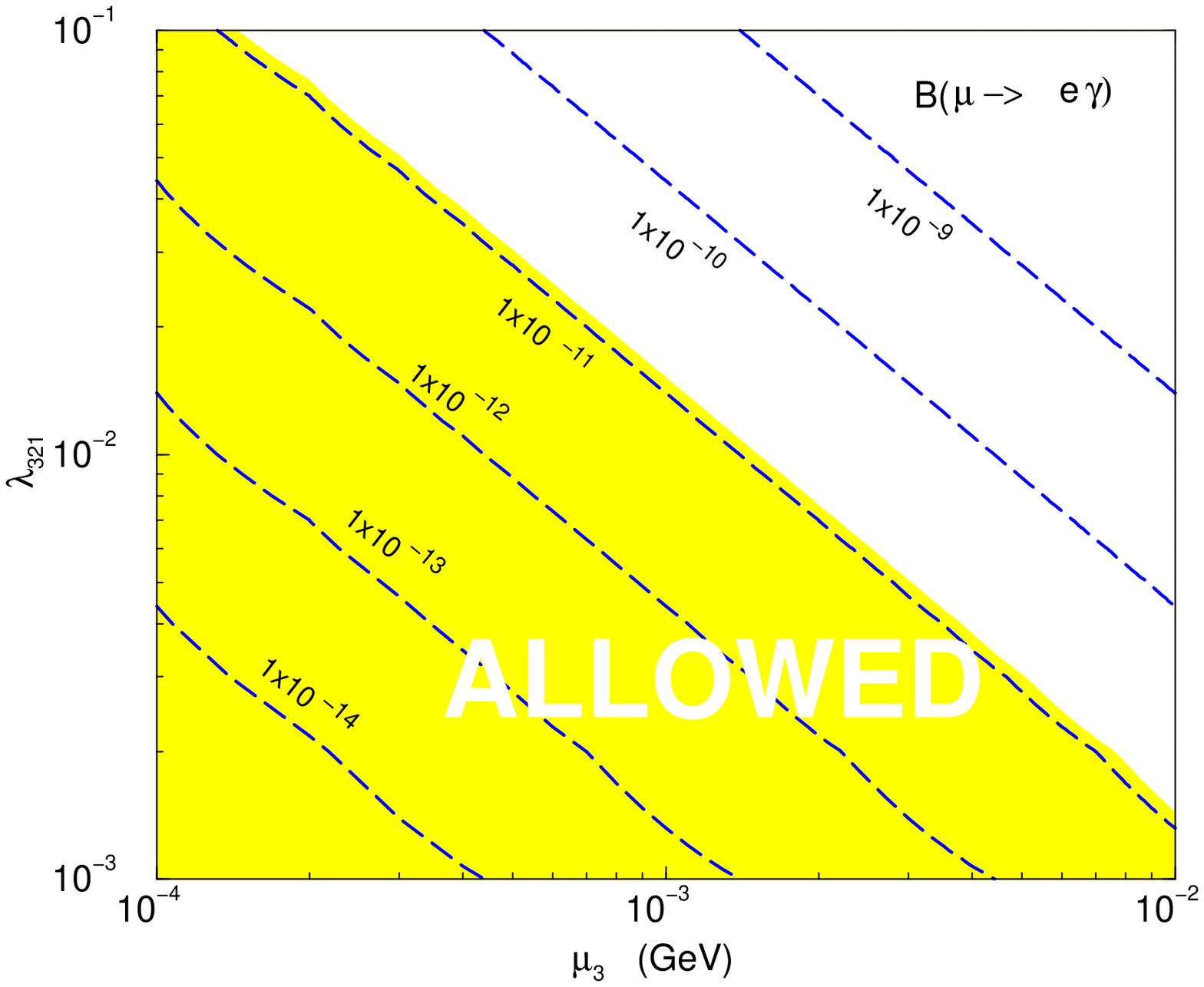}
\caption{\small \label{fig5}
Contours of $B(\mu \to e \,\gamma)$ in the (real) plane of 
$(\mu_{\scriptscriptstyle 3}, \lambda_{\scriptscriptstyle 321})$.
The 90\% C.L. allowed region is shaded.}
\end{figure}

\begin{center} 
\rule{6.5in}{.08mm}
\end{center}

\vspace*{.8in}

\begin{figure}
\vspace*{3.in}
\includegraphics{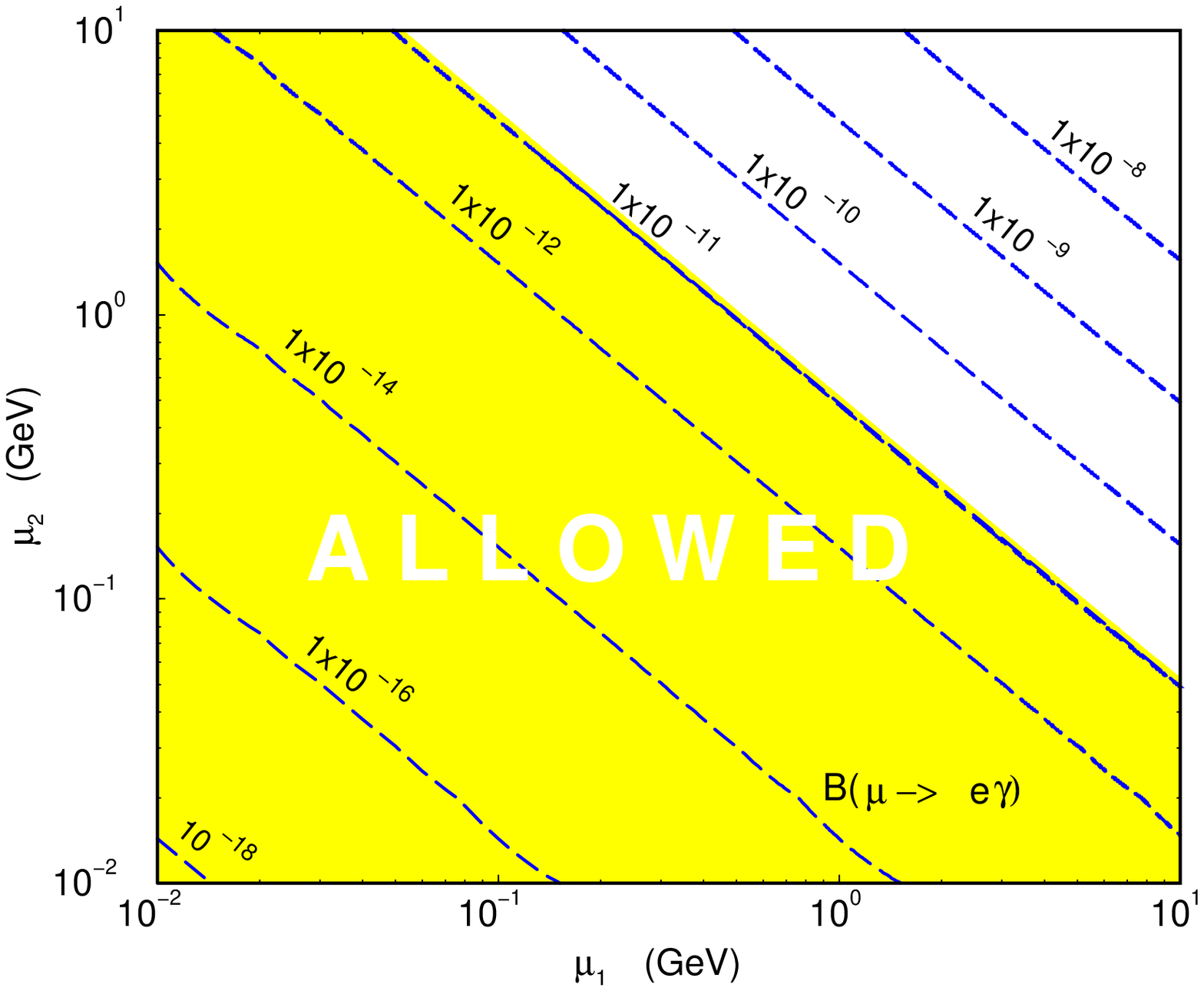}
\caption{\small \label{fig6}
Contours of $B(\mu \to e \,\gamma)$ in the (real) plane of  
(${\mu_{\scriptscriptstyle 1}},{\mu_{\scriptscriptstyle 2}} $).
The 90\% C.L. allowed region is shaded.
Note that the approximation we used for the external lepton
lines is less applicable at the right and top ends of the plot where
the result should be read with caution.}
\end{figure}

\eject

\begin{figure}
\vspace*{3in}
\includegraphics{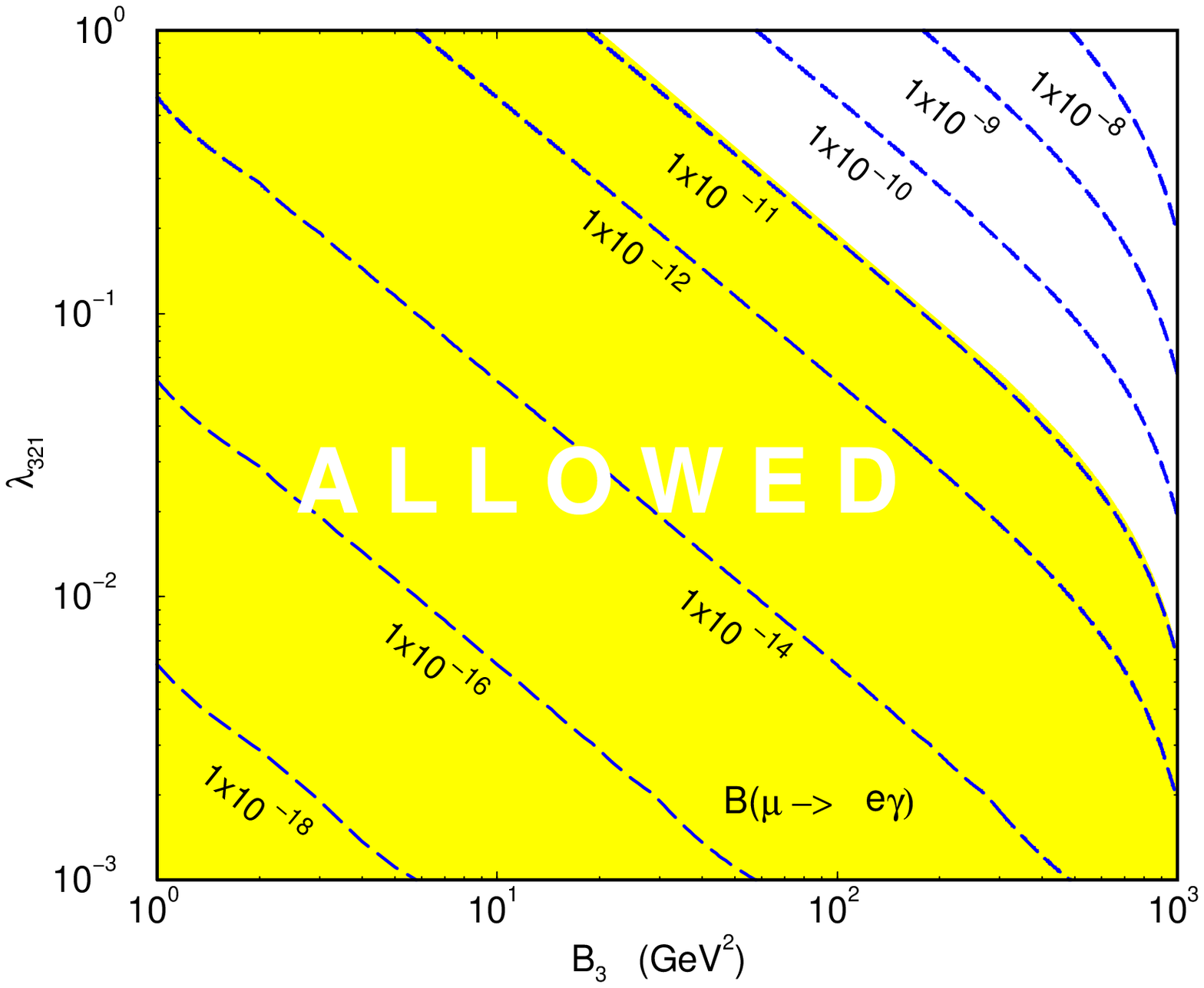}
\caption{\small \label{fig7}
Contours of $B(\mu \to e \,\gamma)$ in the (real) plane of  
(${B_{3}}, \lambda_{\scriptscriptstyle 321}$).
The 90\% C.L. allowed region is shaded. }
\end{figure}

\begin{center} 
\rule{6.5in}{.08mm}
\end{center}

\vspace*{.8in}

\begin{figure}
\vspace*{3.2in}
\includegraphics{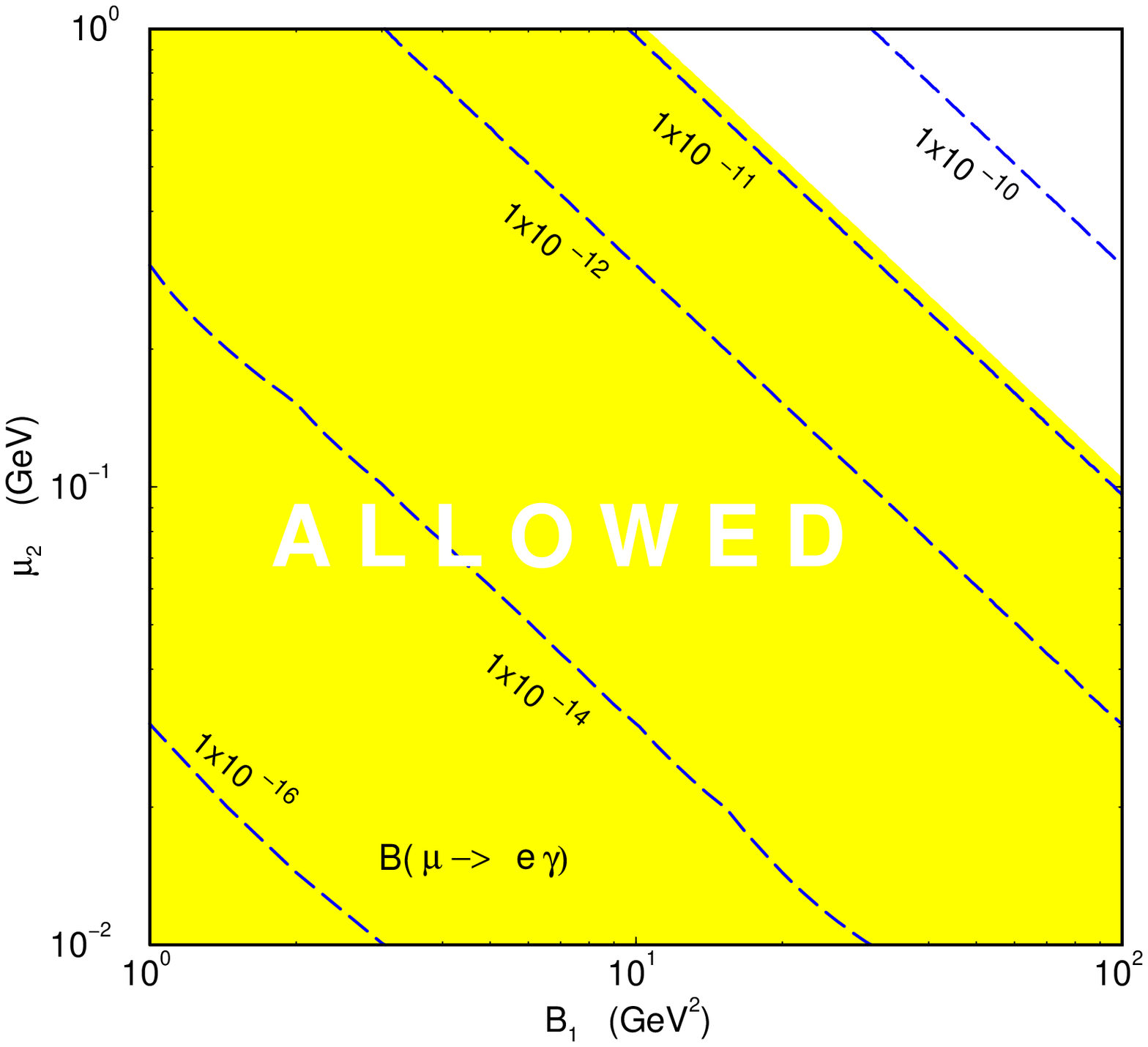}
\caption{\small \label{fig8}
Contours of $B(\mu \to e \,\gamma)$ in the (real) plane of  
($B_1,{\mu_{\scriptscriptstyle 2}} $).
The 90\% C.L. allowed region is shaded.}
\end{figure}


\begin{thebibliography}{99}
\bibitem{smass}
The complete expressions for all scalar masses are first given in Ref.\cite{as5}. However,
the expression corresponding to Eq.(\ref{lastsc}) here was written with a typo --- an extra
numerical factor of 2.
\end{thebibliography}
\end{document}